\documentclass[a4paper]{article}
\usepackage{amsmath}
\usepackage{amsthm}
\usepackage{latexsym}
\usepackage{epsfig}
\usepackage{ifpdf}
\usepackage{color}

\theoremstyle{remark}

\begin{document}

\newcommand{\A}{{\bf A}}
\newcommand{\B}{{\bf B}}
\newcommand{\bco}{{\boldsymbol{:}}}
\newcommand{\blambda}{{\boldsymbol{\lambda}}}
\newcommand{\bmu}{{\boldsymbol{\mu}}}
\newcommand{\bn}{{\bf n}}
\newcommand{\bnabla}{{\boldsymbol{\nabla}}}
\newcommand{\bomega}{{\boldsymbol{\omega}}}
\newcommand{\bsigma}{{\boldsymbol{\sigma}}}
\newcommand{\btheta}{{\boldsymbol{\theta}}}
\newcommand{\bu}{{\bf u}}
\newcommand{\bU}{{\bf U}}
\newcommand{\bv}{{\bf v}}
\newcommand{\bw}{{\bf w}}
\newcommand{\bzero}{{\bf 0}}
\newcommand{\ct}{{\mathcal{T}}}
\newcommand{\cth}{{\mathcal{T}_h}}
\newcommand{\dsum}{{\displaystyle\sum}}
\newcommand{\D}{{\bf D}}
\newcommand{\e}{{\bf e}}
\newcommand{\F}{{\bf F}}
\newcommand{\G}{{\bf G}}
\newcommand{\g}{{\bf g}}
\newcommand{\Gx}{{\stackrel{\rightarrow}{Gx}}}
\newcommand{\I}{{\bf I}}
\newcommand{\intbt}{{\displaystyle{\int_{B(t)}}}}
\newcommand{\intG}{{\displaystyle{\int_{\Gamma}}}}
\newcommand{\into}{{\displaystyle{\int_{\Omega}}}}
\newcommand{\intpb}{{\displaystyle{\int_{\partial B}}}}
\newcommand{\lto}{{L^2(\Omega)}}
\newcommand{\no}{{\noindent}}
\newcommand{\obo}{{\Omega \backslash \overline{B(0)}}}
\newcommand{\obt}{{\overline{B(t)}}}
\newcommand{\oo}{{\overline{\Omega}}}
\newcommand{\R}{{\text{I\!R}}}
\newcommand{\T}{{\bf T}}
\newcommand{\V}{{\bf V}}
\newcommand{\w}{{\bf w}}
\newcommand{\W}{{\bf W}}
\newcommand{\x}{{\bf x}}
\newcommand{\bxi}{{\boldsymbol{\xi}}}
\newcommand{\Y}{{\bf Y}}
\newcommand{\y}{{\bf y}}

\renewcommand{\thefootnote}{\fnsymbol{footnote}}
\footnotetext[1]{Corresponding author. E-mail address: pan@math.uh.edu}

\parskip 4pt
\abovedisplayskip 7pt
\belowdisplayskip 7pt
\parindent=24pt

\begin{center}
{\Large\bf  Circular band formation for incompressible viscous fluid--rigid
particle mixtures in a rotating cylinder}
\vskip 4ex
 Suchung Hou$^1$, Tsorng-Whay Pan$^{2,}$\footnotemark[1] 
 and Roland Glowinski$^2$ 
\vskip 1ex
$^1$Department of Mathematics, National Cheng Kung University, Tainan 701, Taiwan, R.O.C.
\vskip 0.5ex
$^2$Department of Mathematics, University of Houston, Houston, TX 77204, USA
\end{center}
\vskip 8ex
\begin{abstract}
In this paper we have investigated a circular band formation of fluid-rigid
particle mixtures in a fully filled cylinder horizontally rotating about its 
cylinder axis by direct numerical simulation. These phenomena are modeled by 
the Navier-Stokes equations coupled to the Euler-Newton equations describing 
the rigid solid motion of the non-neutrally particles.  The formation of 
circular bands studied in this paper is not resulted by mutual interaction 
between the particles and the periodic inertial waves in the cylinder axis 
direction (as suggested in Phys. Rev. E, 72, 021407 (2005)), but due to the 
interaction of particles.  When a circular band is forming, the part of the 
band formed by the particles moving downward becomes more compact due to the 
particle interaction strengthened by the downward acceleration from the 
gravity. The  part of a band formed by the particles moving upward is always 
loosening up due to the slow down of the particle motion by the counter effect 
of the gravity.  To form a compact circular band (not a loosely one), enough 
particles are needed to interact among themselves continuously through the 
entire circular band at a rotating rate so that the upward diffusion of 
particles can be balanced by the compactness process when these particles 
moving downward. 

\end{abstract}

 
\section{Introduction}

Non-equilibrium systems often organize into interesting spatio-temporal structures or 
patterns. Examples include the patterns in pure fluid flow systems, such as the 
Taylor-Couette flow between two concentric rotating cylinders and  well-defined periodic 
bands of particles in a partially or fully filled horizontally rotating cylinder. 
Particulate flows exhibiting circular bands in a partially filled horizontal 
rotating cylinder are in part attributed to the presence of the free surface caused by 
the partial filling of the cylinder (e.g., see \cite{Acrivos1999,Acrivos2000,Joseph2003}).
In a fully filled horizontally rotating cylinder,  band and other pattern formations 
were also found in the suspensions of non-Brownian settling particles in \cite{Matson2003}-\cite{Seiden2005}.
In this paper, we have focused on the understanding of the band formation closely related to 
those observed in \cite{Lipson2002,Seiden2005}, which are different from those considered
experimentally in \cite{Matson2003, Matson2005, Matson2008, Breu2003} and numerically in 
\cite{Ladd2005, Ladd2007}. The ratio of the particle diameter and the inner cylinder diameter in 
\cite{Lipson2002,Seiden2005} is about 6 to 7\%, which is larger than
the particles of the ratio of 1\%  used in 
\cite{Matson2003, Matson2005, Matson2008, Breu2003,Ladd2005, Ladd2007}.
The cylinder rotating rate is in the range between 6 to 10 sec$^{-1}$ 
in \cite{Lipson2002,Seiden2005} and the fluid-particle mixture is not in the 
Stoke flow regime as considered computationally in \cite{Ladd2005, Ladd2007}.
In \cite{Lipson2001} Lipson used a horizontal rotating cylinder filled with over-saturated 
solution to grow crystal without any interaction with a substrate and found that crystals
accumulate in well-defined periodic bands, normal to the axis of rotation. 
Lipson and Seiden \cite{Lipson2002} just suggested, with no further discussion, that it 
could be the interaction between particles and fluid in the tube. 
In \cite{Seiden2005}, Seiden et al. did an experimental investigation of the dependence of
the formation of bands on particle characteristics, tube diameter and length, and fluid 
viscosity. They suggested that the segregation of particles occurs as a result of mutual 
interaction between the particles and inertial waves excited in the bounded fluid.
In \cite{Seiden2007} Seiden et al.  believed that the axial pressure gradient associated 
with an inertial-mode excitation within bounded fluid is responsible for the formation 
of bands according to their general dimensionless analysis. A single ball motion 
was discussed by solving the equation of motion for the ball with a one-way coupling in a 
filled and horizontally rotating cylinder; a stability analysis and a phase diagram based 
on one ball motion are addressed; but they did not consider the effect of the ball to the 
fluid and the interaction among many particles.

Via the direct numerical simulations, we have found that, at least for all cases considered 
in this paper the formation of circular bands is not resulted by mutual interaction between the 
particles and the periodic inertial waves in the cylinder axis direction as suggested 
in \cite{Seiden2005,Seiden2007}, but by the interaction between particles. In our simulations,
the particles form a layer inside a horizontally rotating cylinder as in Figure 7 
in \cite{Seiden2005}. These particles are partially coated on the inner wall 
of the rotating cylinder under the influence of a strong centrifugal force. When a circular 
band is forming, the part of the band formed by the particles moving downward becomes more compact 
due to the particle interaction strengthened by the downward acceleration from 
the gravity. The  part of a band formed by the particles moving upward is always 
loosening up due to the slow down of the particle motion by the counter effect 
of the gravity.  To form a compact circular band (not a loosely one), enough particles 
are needed to interact among themselves continuously through the entire circular band at a rotating rate
so that the compactness of the whole circular band can be maintained when these 
particles moving upward.  Hence the balance of the gravity, the rotating rate, and 
the fluid flow inertia plus enough number of particles is important on the formation 
of circular bands in a fully filled cylinder. 

The scheme of this paper is as follows: We discuss the models and numerical methods 
briefly in Sec. 2. In Sec. 3, we study the effect of the particle number and rotating 
rate  on the formation of circular bands and then present
the flow field development under the influence of the particle interaction to
show that the formation of circular bands studied in this paper 
is not resulted by the periodic inertial waves in the cylinder axis direction.
The conclusions are summarized in Sec. 4.


\section{Model and numerical method}

\begin{figure}[ht]
\begin{center}
\leavevmode
\epsfxsize=2.5in
\epsffile{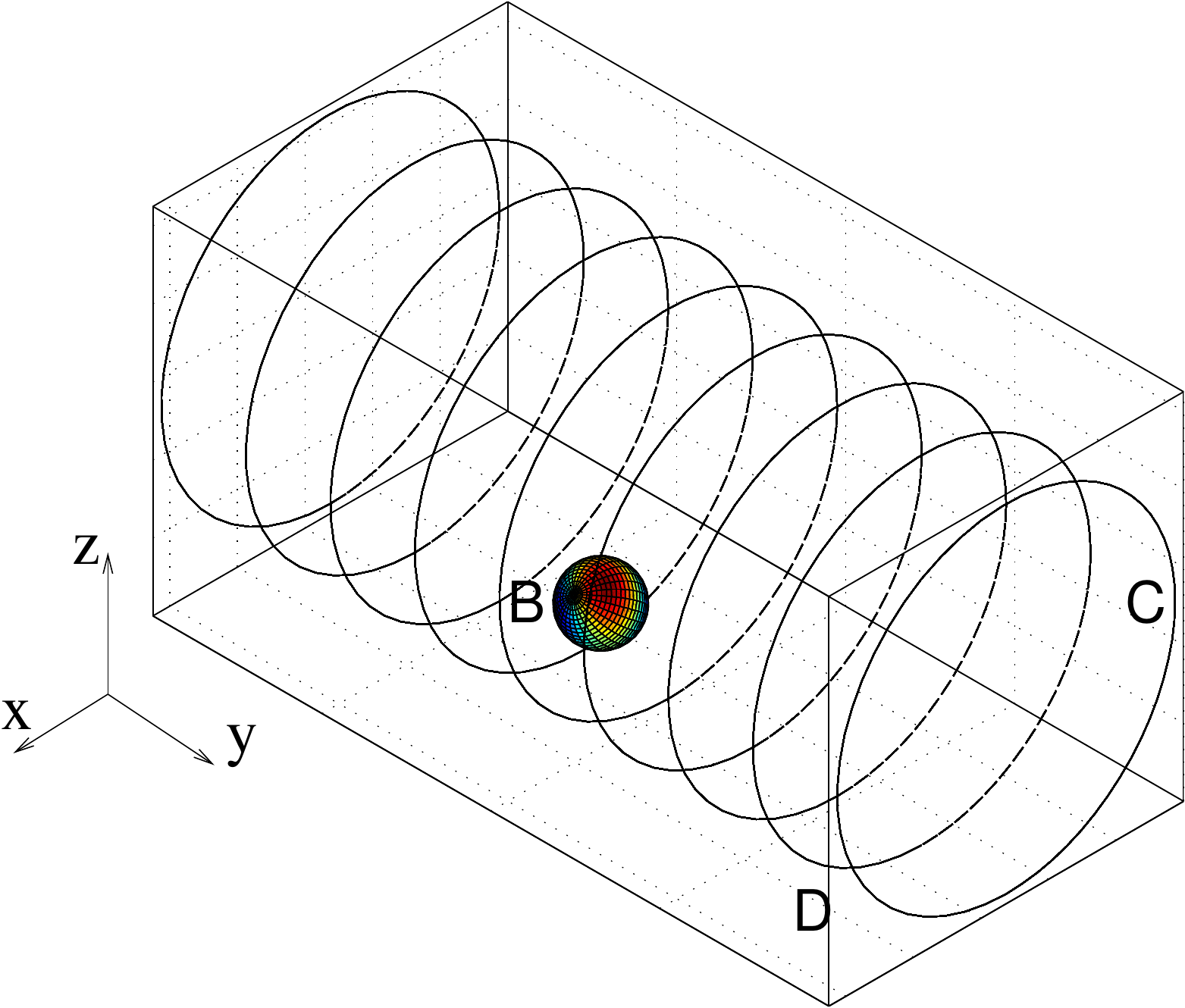}
\epsfxsize=0.2in
\epsffile{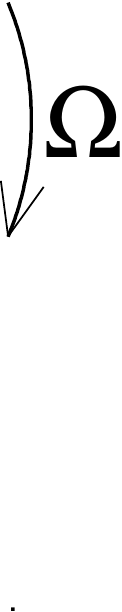}
\end{center}
\caption{The flow region with a ball $B$ in a truncated cylinder $\bf C$.}\label{fig.1}
\end{figure}

To perform the {\it direct numerical simulation} of the interaction between rigid bodies 
and fluid, we have developed a methodology which combines a distributed Lagrange multiplier based
fictitious domain  method with operator splitting and finite element methods (e.g., see
\cite{Pan1999}-\cite{Pan2007}).  
For a ball $B$ moving in a Newtonian viscous incompressible 
fluid of the viscosity $\mu$ and the density $\rho$  contained in a truncated 
cylinder $\bf C$ under the effect of the gravity depicted in Figure \ref{fig.1}, 
the flow is modeled by the Navier-Stokes equations, namely,

\begin{equation}
\rho \Big[\dfrac{\partial {\bf u}}{\partial t}+({\bf u} \cdot
{\boldsymbol{\nabla}}){\bf u}\Big] - \mu \Delta {\bf u}
+ {\boldsymbol{\nabla}} p =  {\bf g} 
\ \, in \ \, \{(\x,t)| \x \in {\bf C} \setminus \overline{B(t)}, \ t \in (0,T)\},
\label{eqn:rg1}
\end{equation}
\begin{equation}
{\boldsymbol{\nabla}} \cdot {\bf u}(t)=0 \ \ in \ \, \{(\x,t)| \x \in {\bf C} \setminus \overline{B(t)}, \ t \in (0,T)\} ,
\label{eqn:rg2}
\end{equation}
\begin{equation}
{\bf u}(0)={\bf u}_0 ({\bf x}), \ (with \ {\boldsymbol{\nabla}} \cdot {\bf u}_0 =0),
\label{eqn:rg3}
\end{equation}
\begin{equation}
{\bf u}= {\bf g}_0 \ on \ \Gamma_0 \times (0,T), \ (with \
\displaystyle\int_{\Gamma_0} {\bf g}_0 \cdot {\bf n} \,d \Gamma =0), 
\label{eqn:rg4}
\end{equation}
where $\Gamma_0$ is the boundary of a truncated cylinder ${\bf C}$, $\g$ denotes gravity and 
$\bn$ is the unit normal vector pointing outward to the flow region. 
We assume a {\it no-slip condition} on $\gamma(=\partial B)$. 
The motion of the rigid body $B$ satisfies the Euler-Newton's equations, namely
\begin{equation}
{\bf v}({\bf x}, t)={\bf V}(t) + \overrightarrow{{\boldsymbol{\omega}}}(t) \times
\overrightarrow{{\bf G}(t){\bf x}}, \ \forall {\bf x} \in \obt, \ \forall t \in (0, T), \label{eqn:rg5}
\end{equation}
\begin{equation}
\dfrac{d {\bf G}}{d t}={\bf V},
\label{eqn:rg6}
\end{equation}
\begin{equation}
M_p \dfrac{d \bf V}{d t} = M_p \, {\bf g}+{\bf F}_H,
\label{eqn:rg7}
\end{equation}
\begin{equation}
\I_p \dfrac{d {\boldsymbol{\omega}}}{dt} ={\bf T}_H,
\label{eqn:rg8}
\end{equation}
with the resultant and torque of the hydrodynamical forces given by,
respectively,
\begin{equation}
{\bf F}_H=-\displaystyle\int_\gamma \boldsymbol{\sigma} {\bf n}\, d
\gamma, \ \ {\bf T}_H=-\displaystyle\int_\gamma \overrightarrow{{\bf Gx}} \times
\boldsymbol{\sigma} {\bf n} \,d \gamma
\label{eqn:rg9}
\end{equation}
with $\boldsymbol{\sigma}=  \mu ({\boldsymbol{\nabla}}{\bf u}+{\boldsymbol{\nabla}}{\bf u}^t)-p \I$.
Equations (\ref{eqn:rg1})--(\ref{eqn:rg9}) are
completed by the following initial conditions
\begin{equation}
{\bf G}(0)={\bf G}_0, \ {\bf V}(0)={\bf V}_0, \ {\boldsymbol{\omega}}(0)
= {\boldsymbol{\omega}}_0,   B(0)=B_0.
\label{eqn:rg10}
\end{equation}
Above, $M_p$, $\I_p$, $\bf G$, $\bf V$ and ${\boldsymbol{\omega}}$ are the mass,
inertia, center of mass, velocity of the center of mass and angular velocity of
the rigid body $B$, respectively. The gravity is pointed downward in the direction of $z$.

To solve numerically the coupled problem (\ref{eqn:rg1})-(\ref{eqn:rg10}), we have applied 
a distributed Lagrange multiplier-based fictitious domain method. Its basic idea is to 
imagine that the fluid fills the space inside as well as outside the particle boundaries. 
The fluid flow problem is then posed on a larger domain $\bf D$, the ``fictitious 
domain''. The fictitious domain has a simple shape, allowing a simple regular mesh to be used. 
This domain is also time-independent, so the same mesh can be used for the entire simulation. 
This is a great advantage, since for simulating 3D interaction of fluid and particles, the 
automatic generation of unstructured boundary-fitted meshes for a large number of closely spaced 
particles considered in this paper is a difficult problem. The fluid inside the particle boundary 
must exhibit a rigid body motion of the particle. This constraint is
enforced using a distributed Lagrange multiplier, which represents the additional body force per
unit volume needed to maintain the rigid body motion inside the particle boundary, much like the
pressure in incompressible fluid flow, whose gradient is the force required to maintain the 
constraint of incompressibility.  For space discretization, we have used $P_1$-{\it iso}-$P_2$ 
and $P_1$ finite elements for the velocity field and pressure, respectively 
(e.g., see \cite{Glowinski2003, Pan2007}). In time advancing, via the
Lie's scheme \cite{Chorin1978} with the finite element approximation, the distributed Lagrange multiplier-based fictitious domain formulation of problem (\ref{eqn:rg1})-(\ref{eqn:rg10}) 
is decoupled into a sequence of simpler subproblems at each time step and solved numerically
(please see \cite{Pan2007} for details).

\section{Numerical experiments and discussion}

\subsection{The combined effect of the rotating rate and the number of particles}
\begin{figure}
\begin{center}
\leavevmode
\hskip 0.36in
\epsfxsize=1.1in
\epsffile{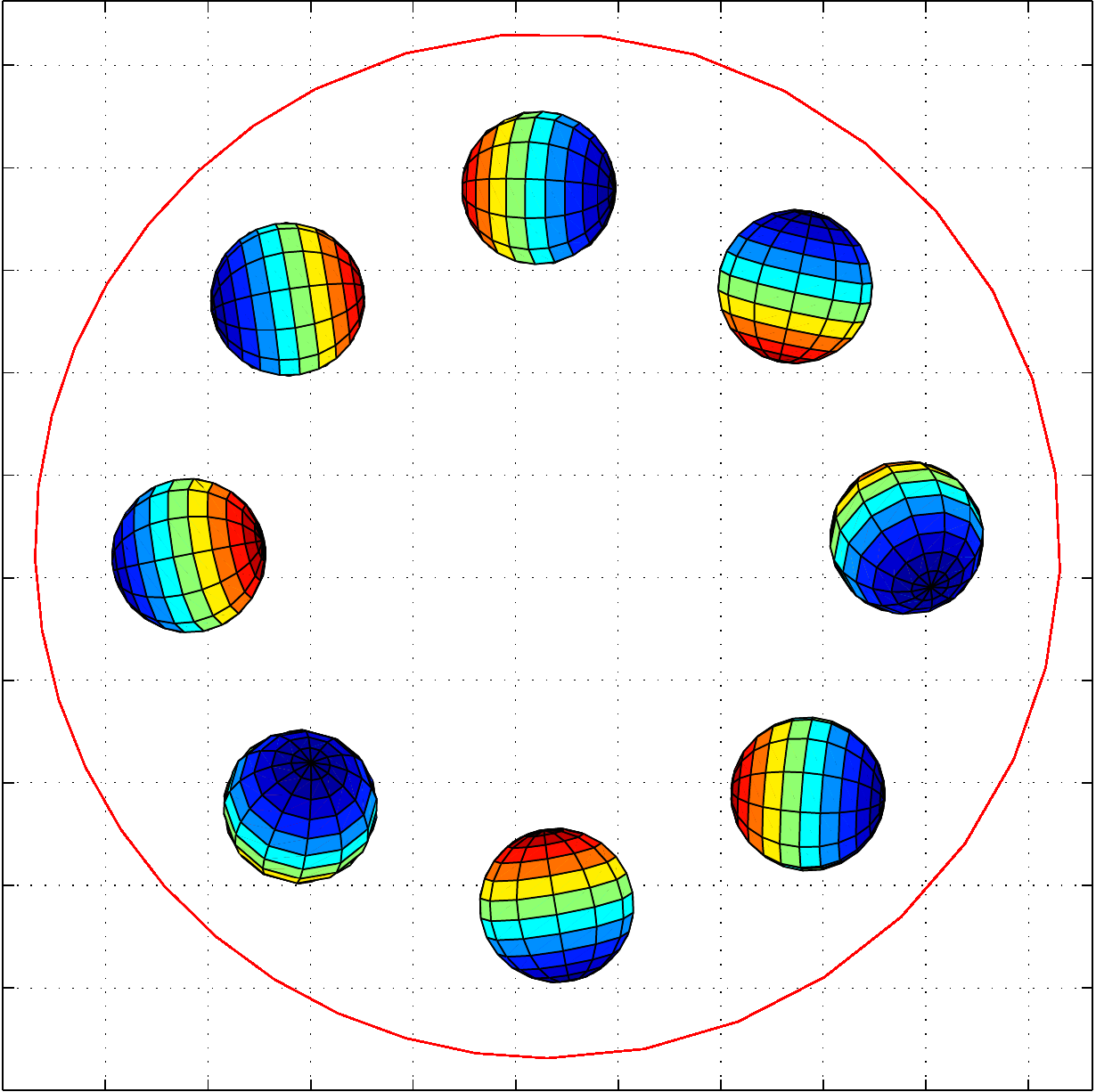}
\hskip 0.37in
\epsfxsize=3.8in
\epsffile{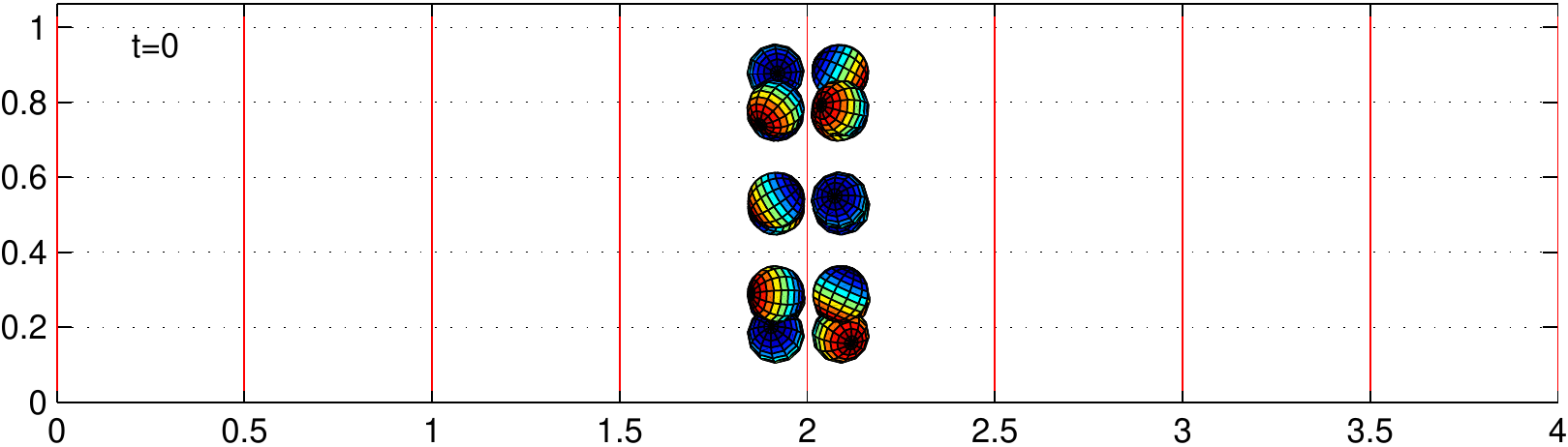}\\ 
\hskip 0.1in
\epsfxsize=0.2in
\epsffile{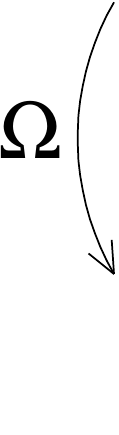}
\epsfxsize=1.1in
\epsffile{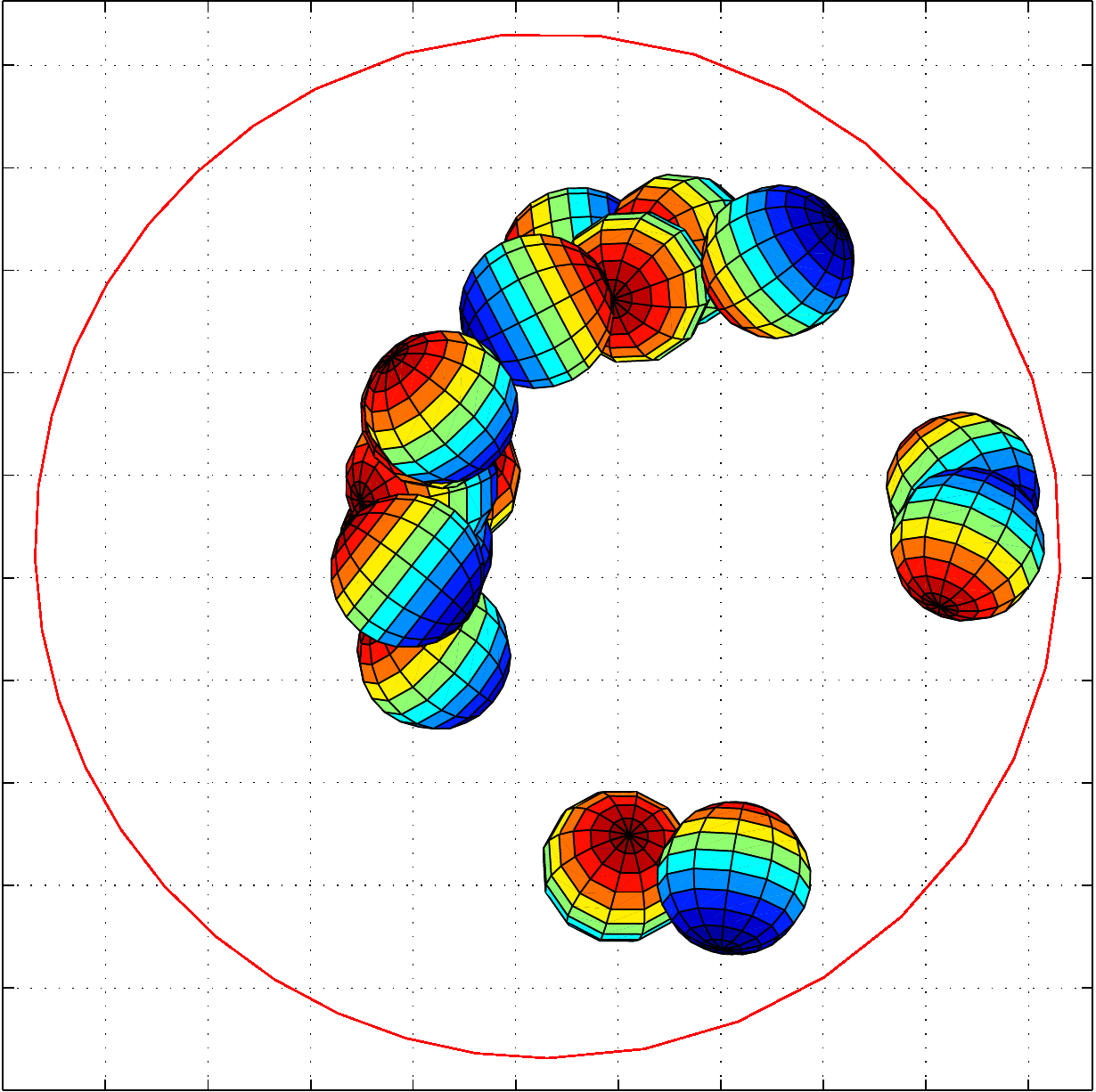} 
\hskip 0.37in
\epsfxsize=3.8in
\epsffile{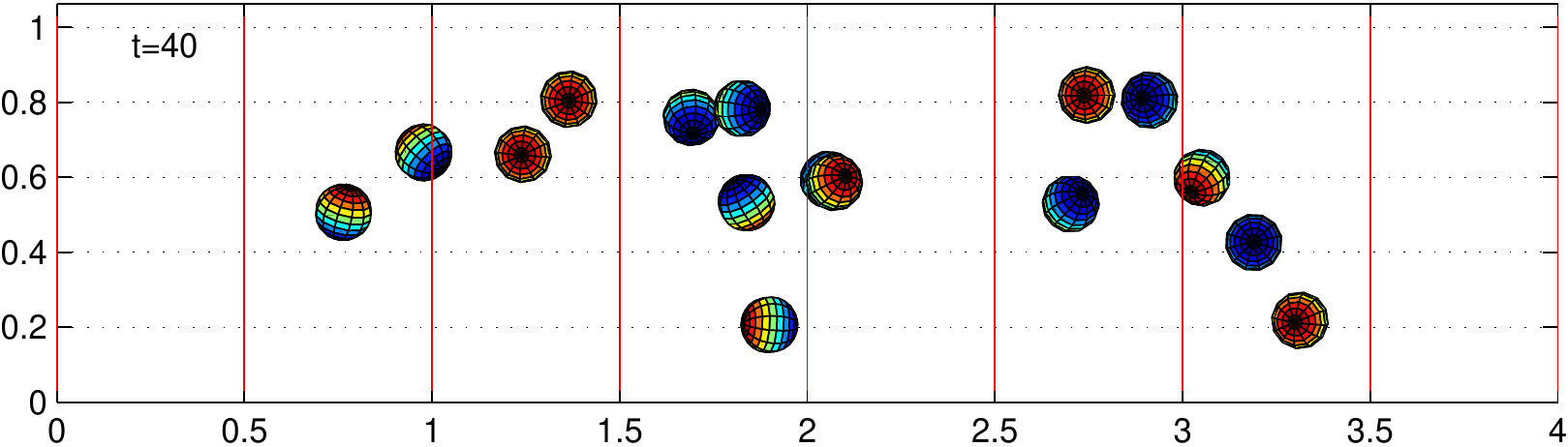}\\
\epsfxsize=0.3in
\epsffile{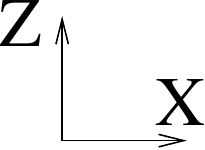}
\epsfxsize=1.1in
\epsffile{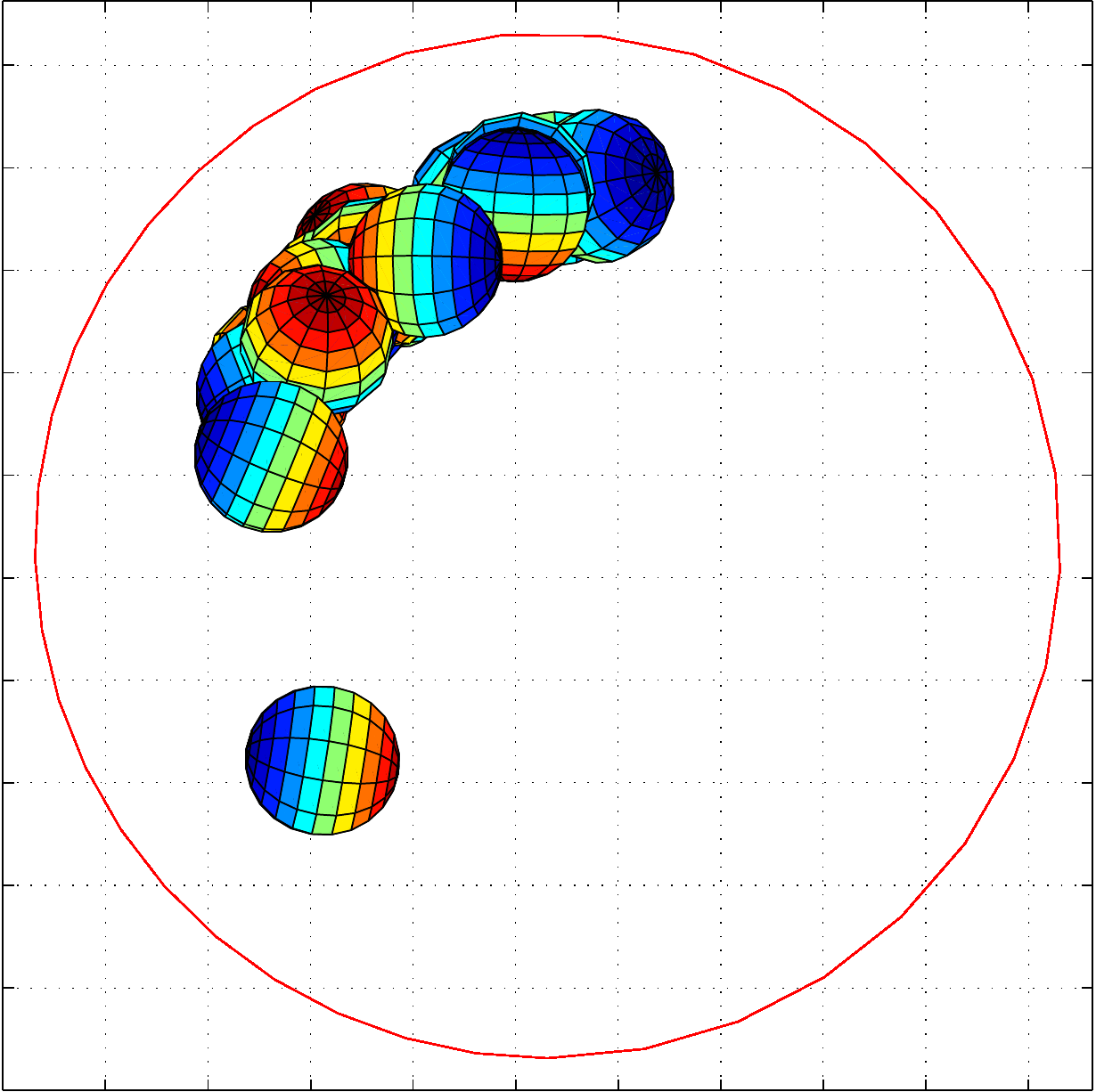}
\epsfxsize=0.3in
\epsffile{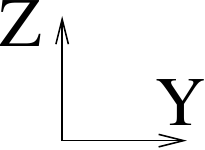}
\epsfxsize=3.8in
\epsffile{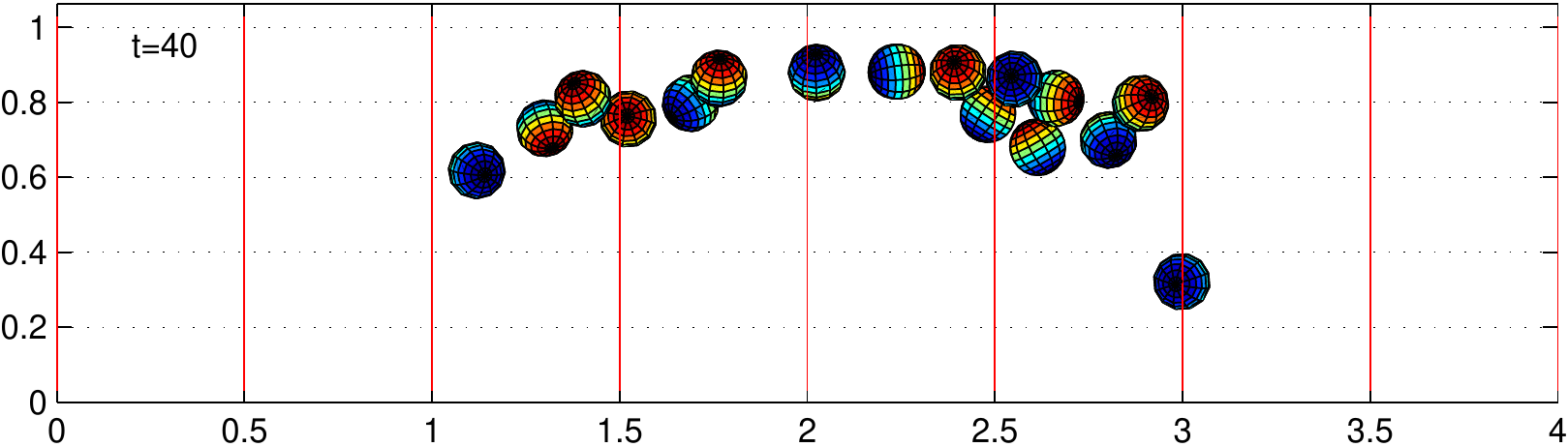}
\end{center}
\caption{The side view (left) and the front view (right) of the 16 ball initial position (top) and the position obtained at the rotating rate $\Omega=$ 8 (middle) and 12 (bottom) sec$^{-1}$ at $t=$ 40 second.}\label{fig.3}
\begin{center}
\leavevmode
\hskip 0.36in
\epsfxsize=1.1in
\hskip 1.57in
\epsfxsize=3.8in
\epsffile{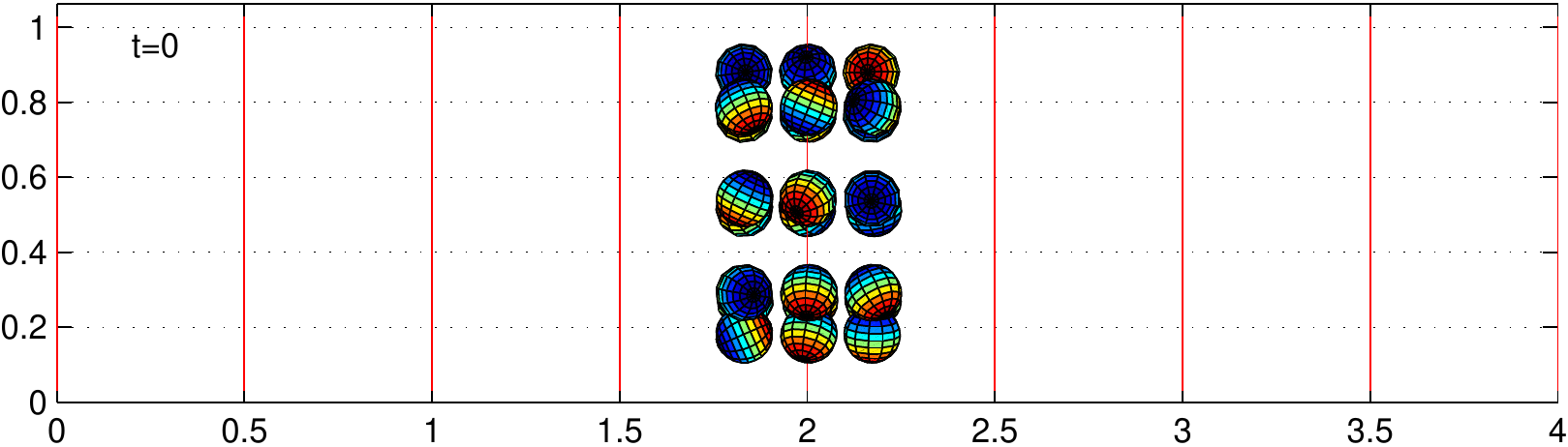}\\  
\hskip 0.1in
\epsfxsize=0.2in
\epsffile{omega.pdf}
\epsfxsize=1.1in
\epsffile{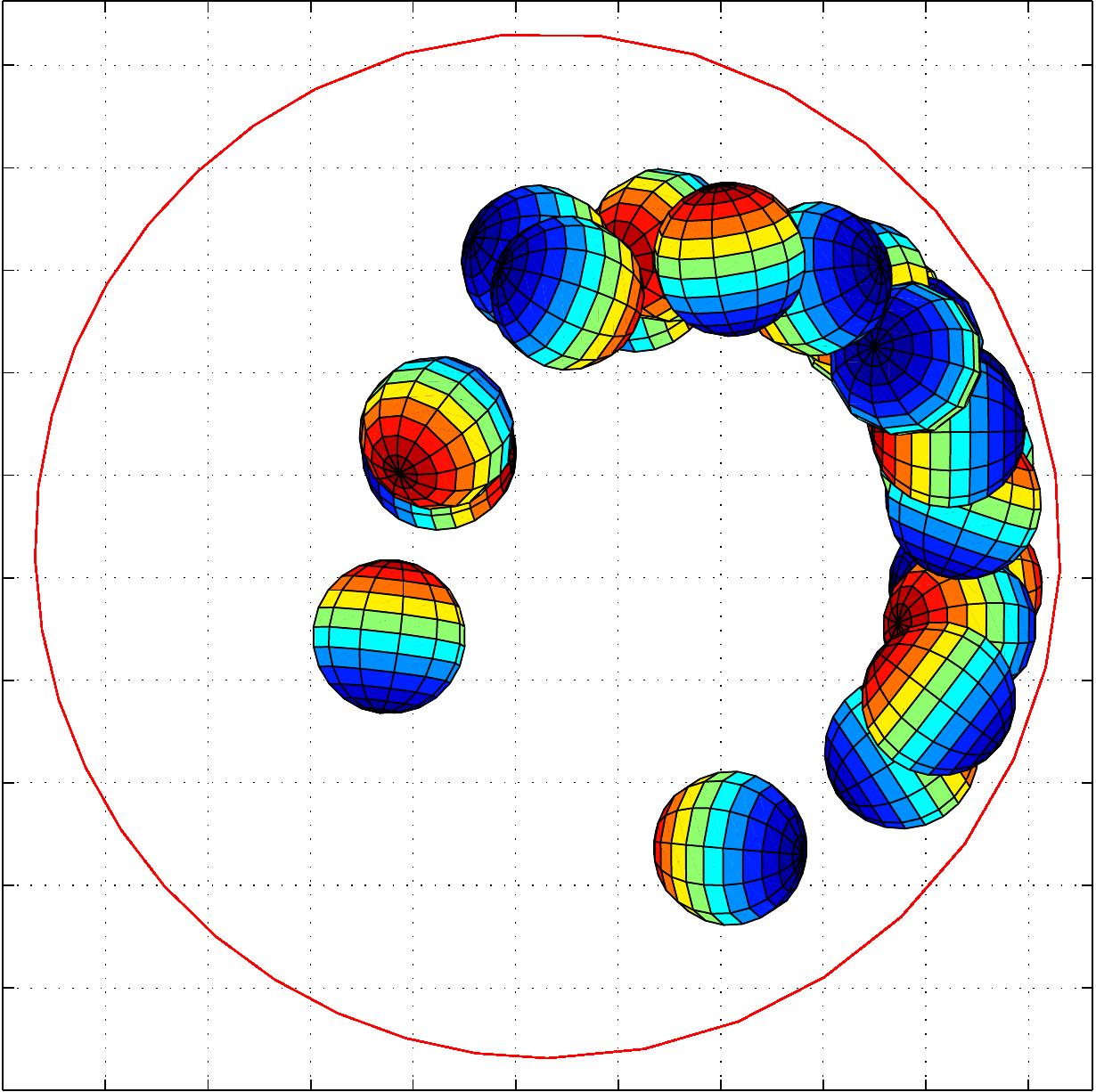} 
\hskip 0.37in
\epsfxsize=3.8in
\epsffile{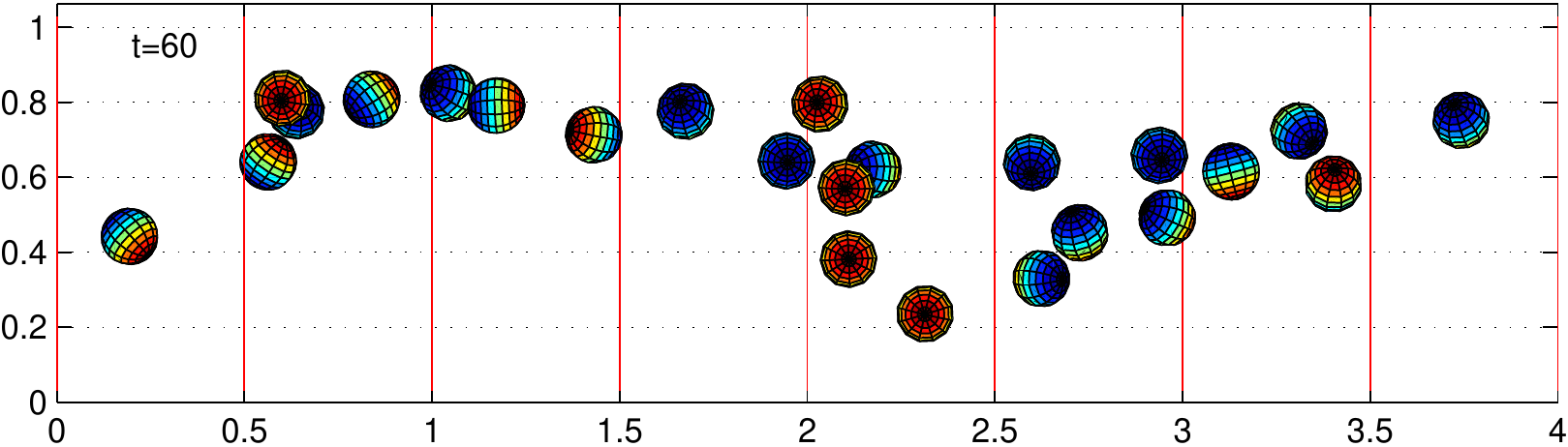}\\  
\epsfxsize=0.3in
\epsffile{xz-coor.pdf}
\epsfxsize=1.1in
\epsffile{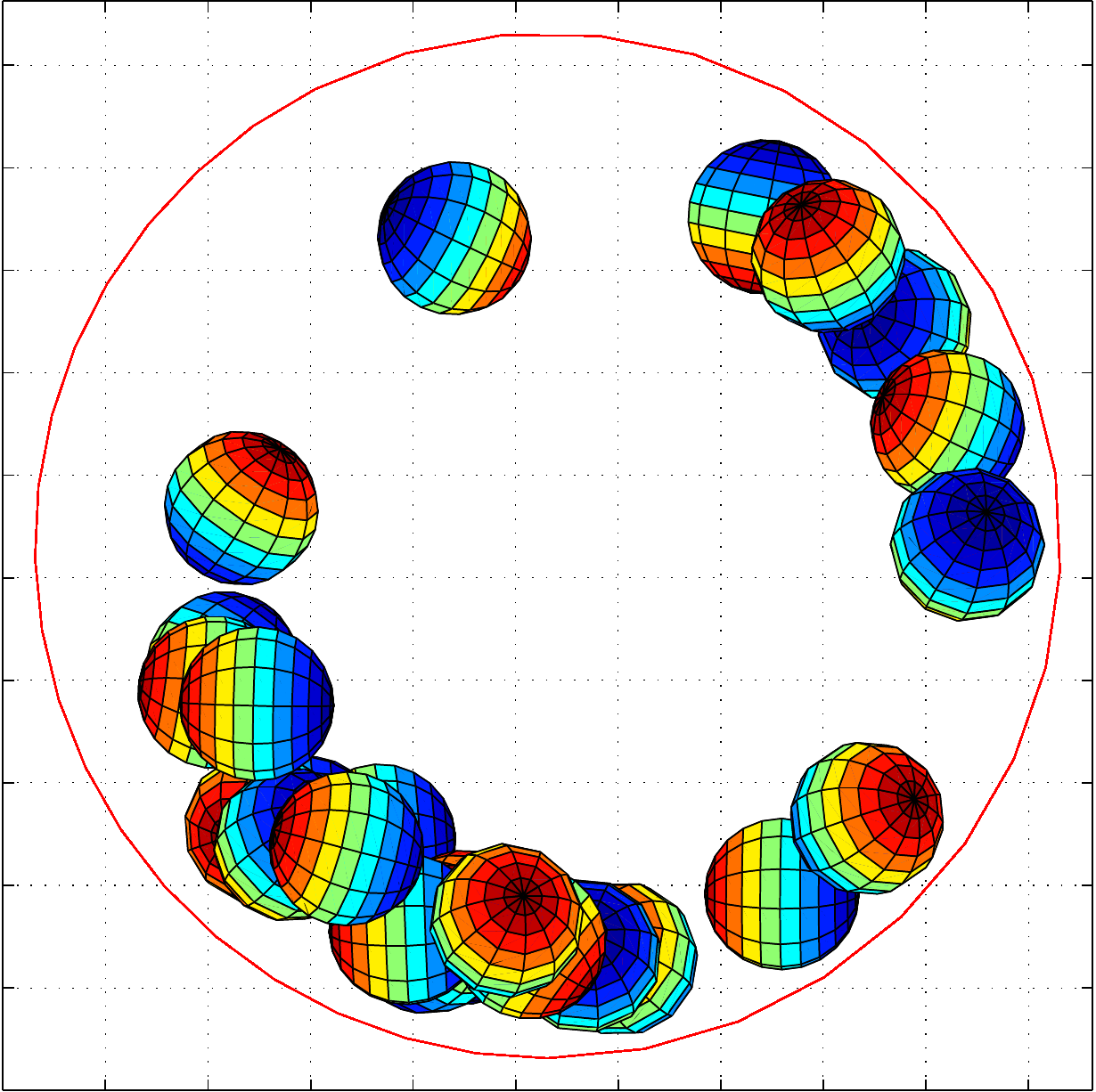}
\epsfxsize=0.3in
\epsffile{yz-coor.pdf}
\epsfxsize=3.8in
\epsffile{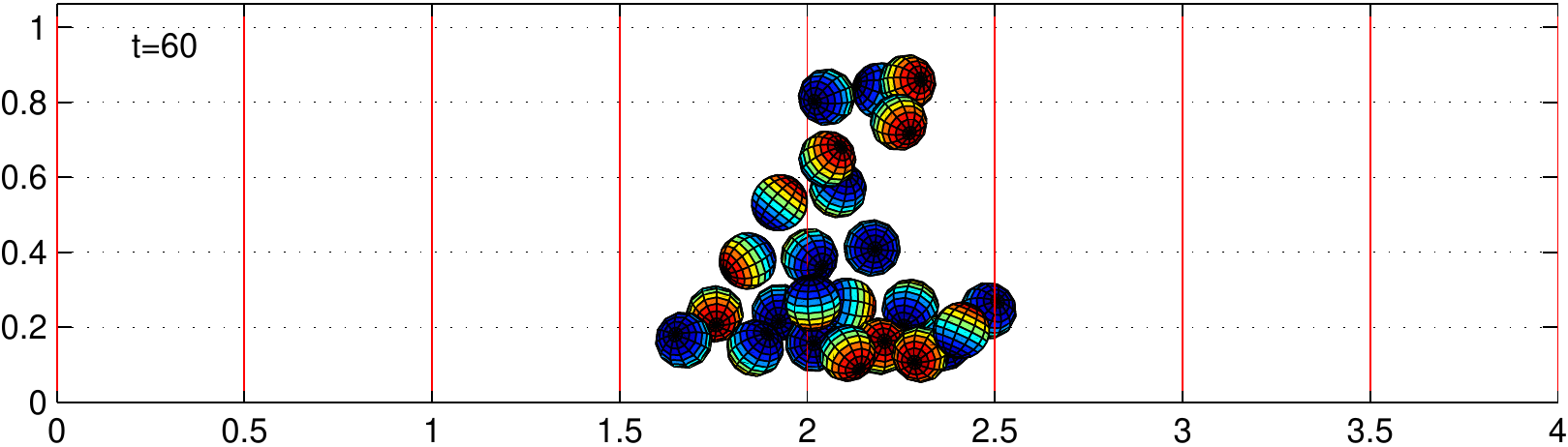}
\end{center}
\caption{The side view (left) and the front view (right) of the 24 ball initial position (top) and the position obtained at the rotating rate $\Omega=$ 8 (middle) and 12 (bottom) sec$^{-1}$ at $t=$ 60 second.}\label{fig.4}
\end{figure}

\begin{figure}
\begin{center}
\leavevmode
\hskip 0.36in
\epsfxsize=1.1in
\hskip 1.57in
\epsfxsize=3.8in
\epsffile{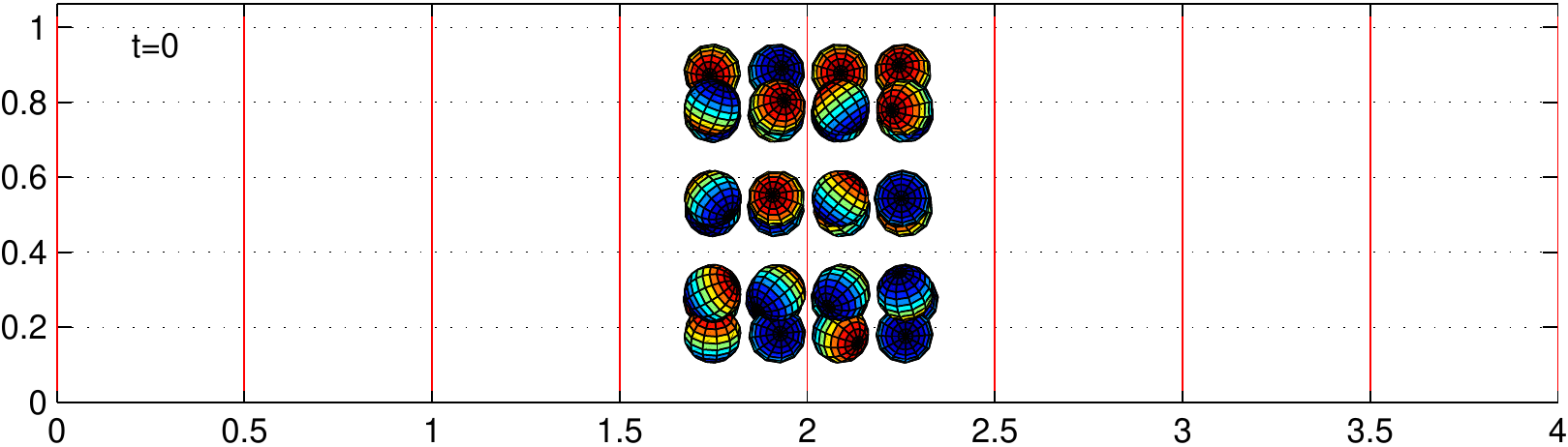}\\ 
\hskip 0.1in
\epsfxsize=0.2in
\epsffile{omega.pdf}
\epsfxsize=1.1in
\epsffile{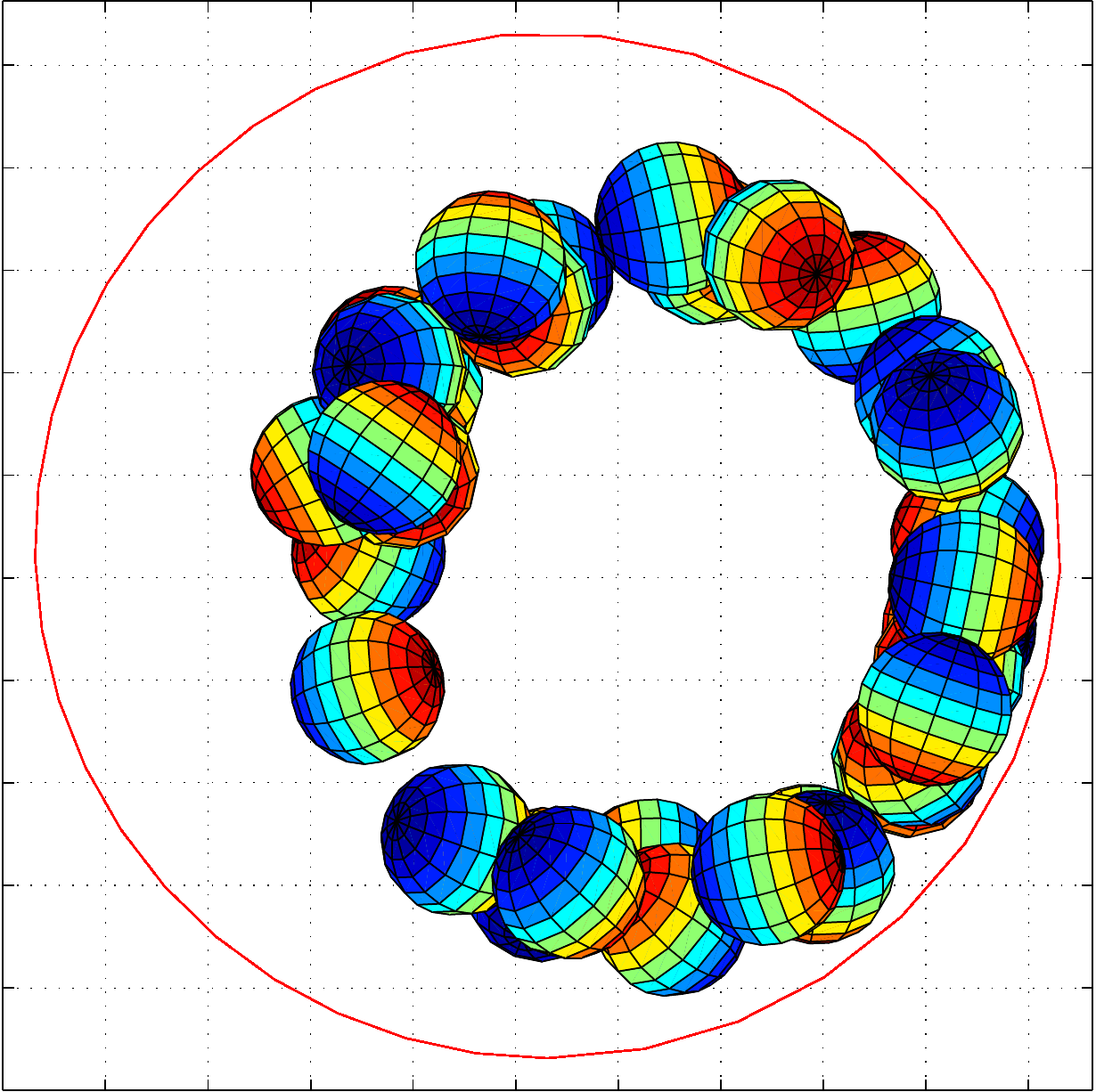}
\hskip 0.37in
\epsfxsize=3.8in
\epsffile{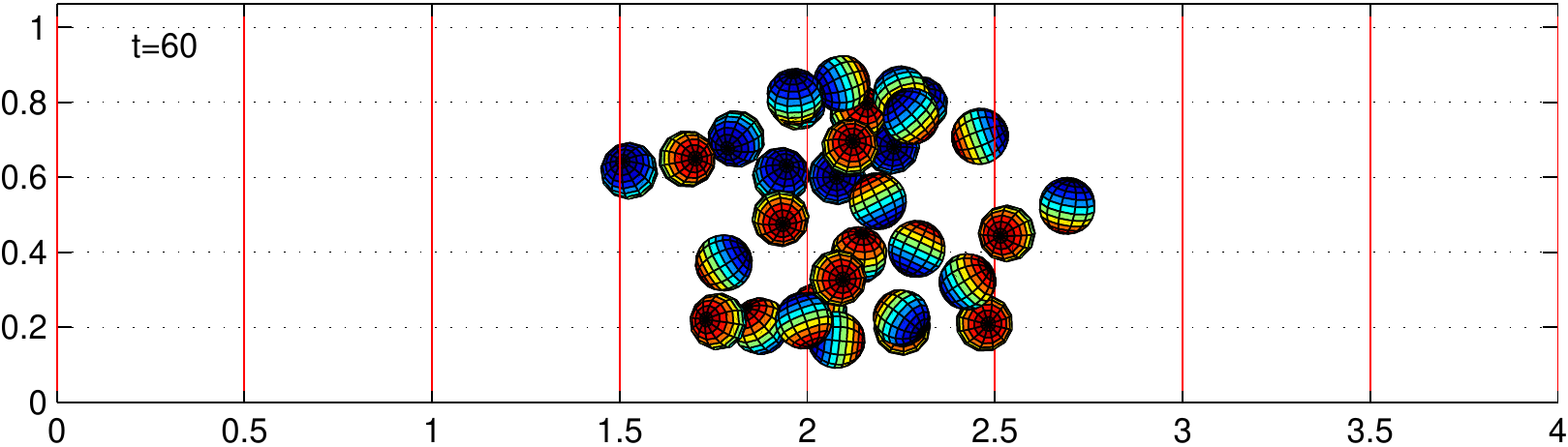}\\
\epsfxsize=0.3in
\epsffile{xz-coor.pdf}
\epsfxsize=1.1in
\epsffile{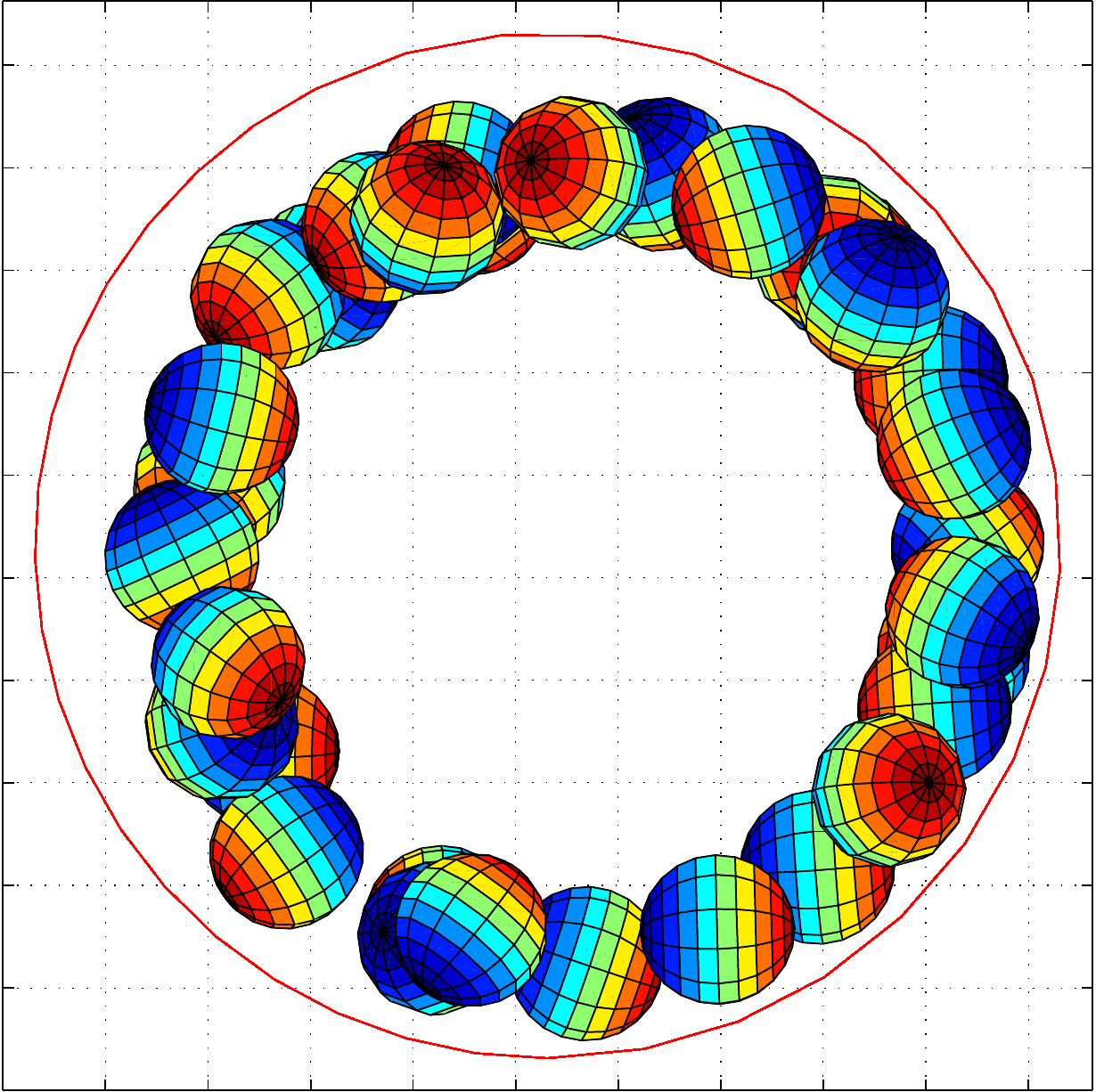}
\epsfxsize=0.3in
\epsffile{yz-coor.pdf}
\epsfxsize=3.8in
\epsffile{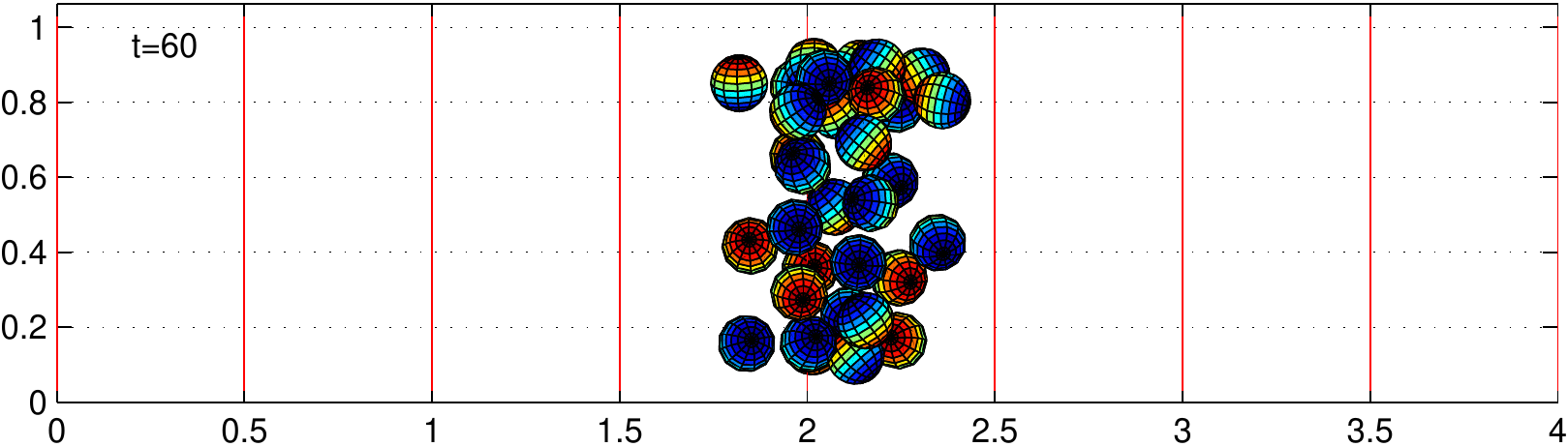}
\end{center}
\caption{The side view (left) and the front view (right) of the 32 ball initial position (top) and the position obtained at the rotating rate $\Omega=$ 8 (middle) and 12 (bottom) sec$^{-1}$ at $t=$ 60 second.}\label{fig.5}
\begin{center}
\leavevmode
\hskip 0.36in
\epsfxsize=1.1in
\hskip 1.57in
\epsfxsize=3.9in
\epsffile{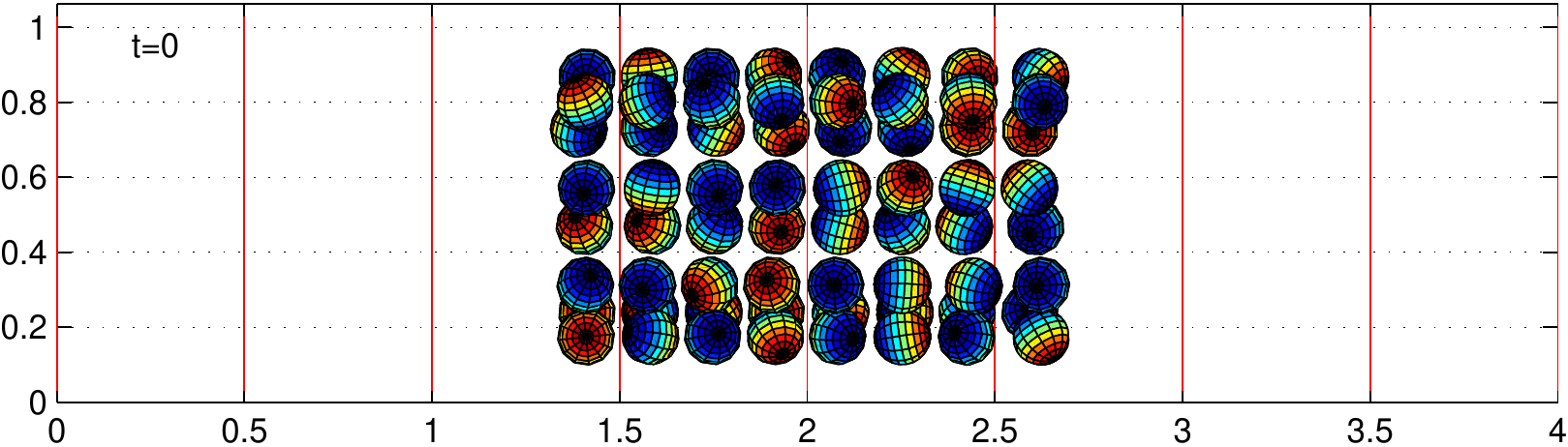}\\  
\hskip 0.1in
\epsfxsize=0.2in
\epsffile{omega.pdf}
\epsfxsize=1.1in
\epsffile{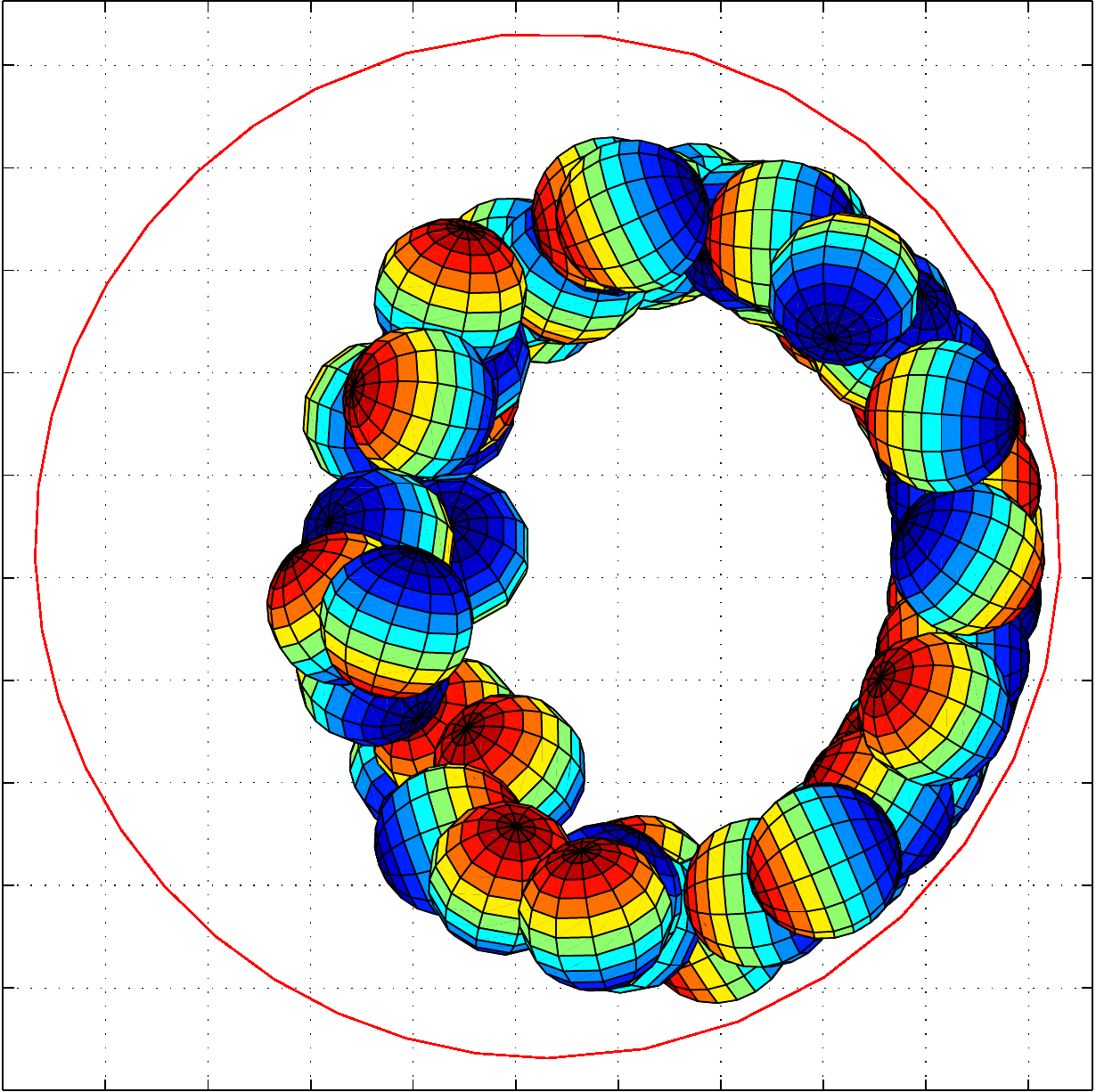} 
\hskip 0.37in
\epsfxsize=3.9in
\epsffile{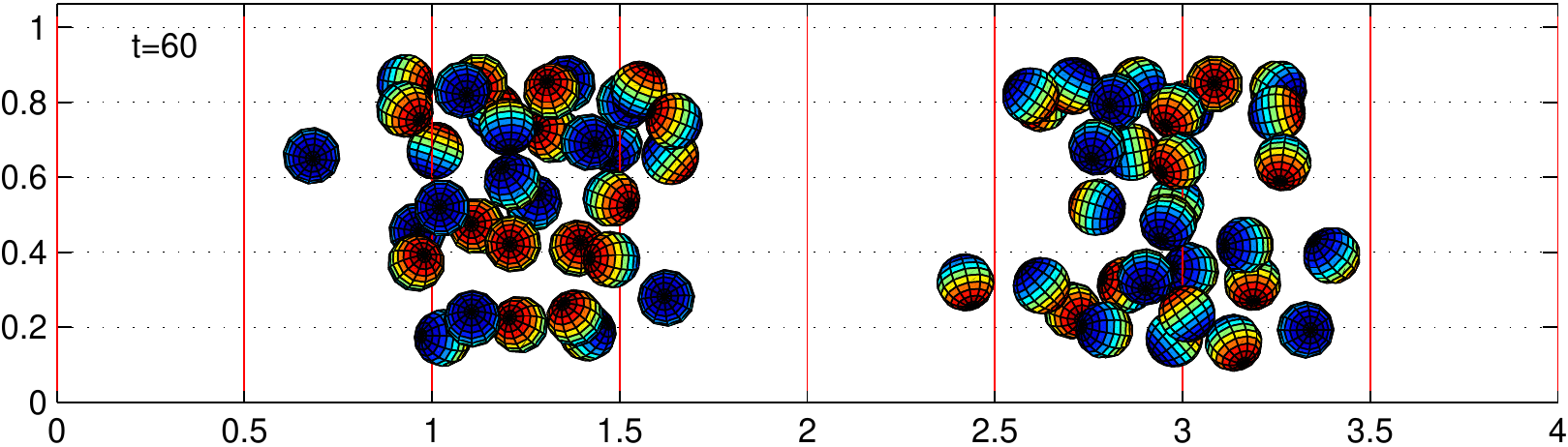}\\ 
\epsfxsize=0.3in
\epsffile{xz-coor.pdf}
\epsfxsize=1.1in
\epsffile{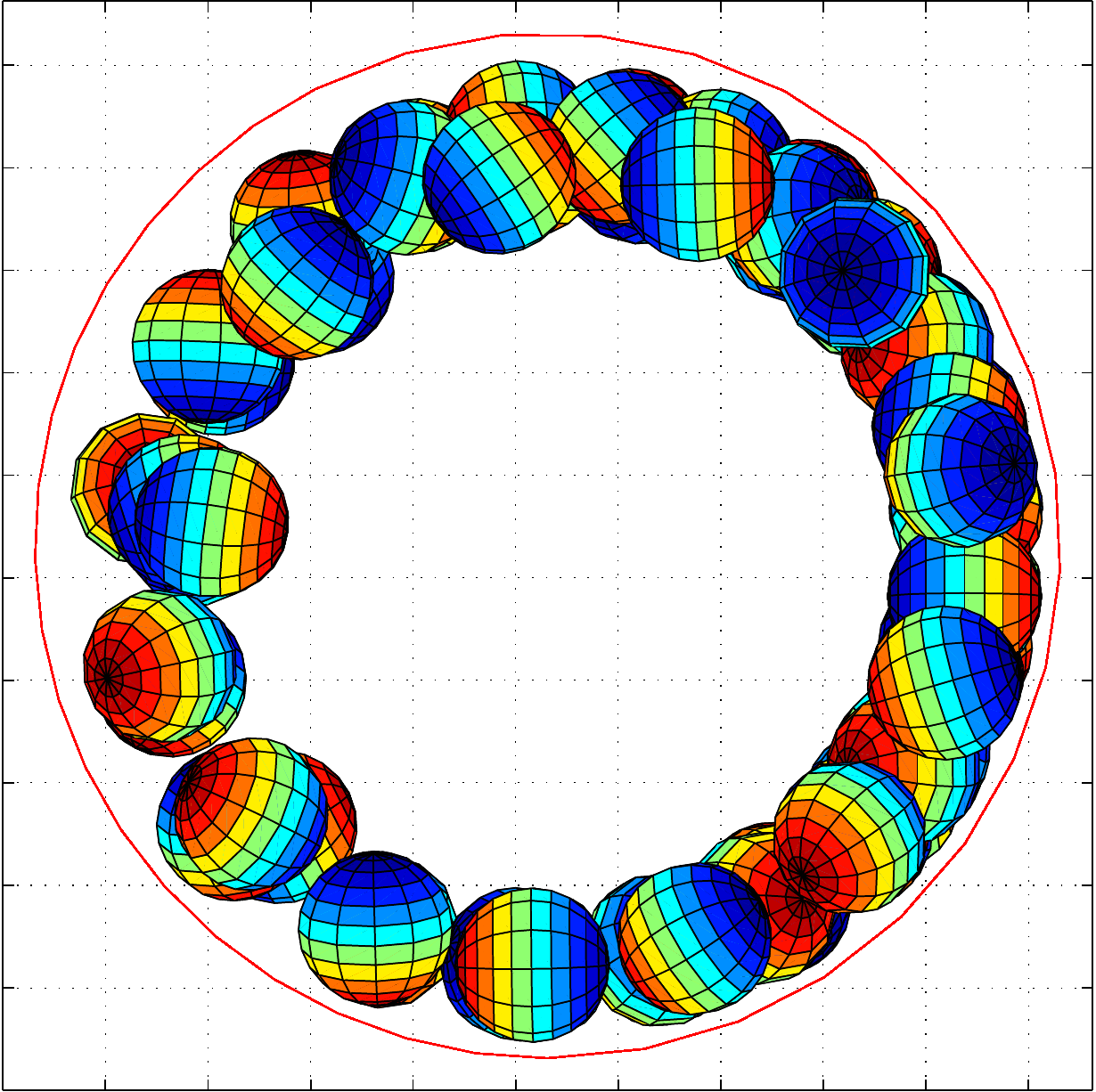} 
\epsfxsize=0.3in
\epsffile{yz-coor.pdf}
\epsfxsize=3.9in
\epsffile{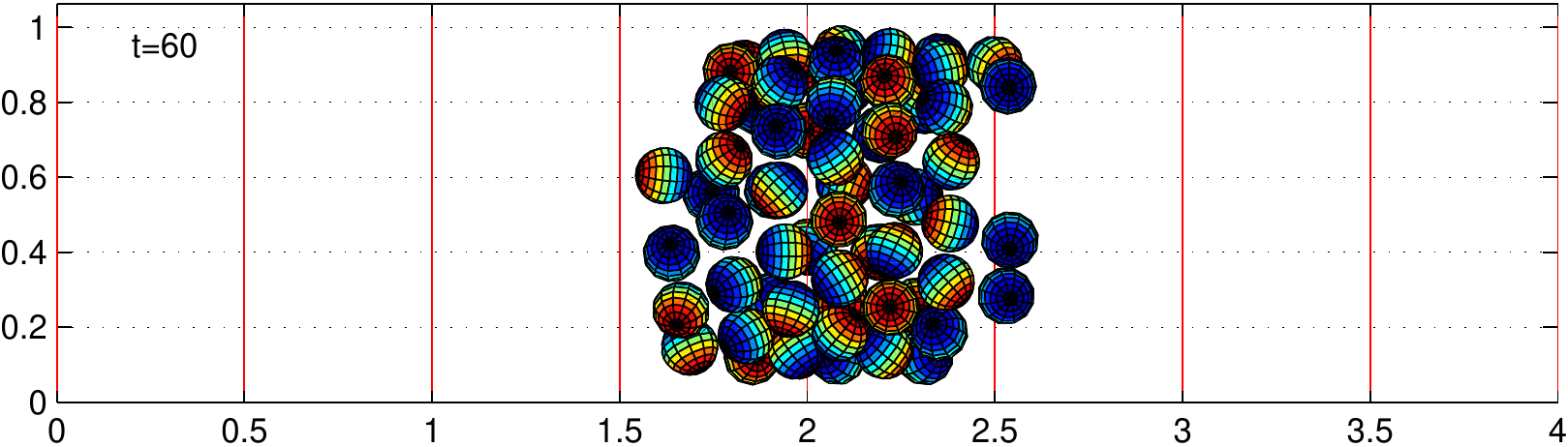}
\end{center}
\caption{The side view (left) and the front view (right) of the  64 ball initial position (top) and the position obtained at the rotating rate $\Omega=$ 8 (middle) and 12 (bottom) sec$^{-1}$ at $t=$ 60 second.}\label{fig.6}
\end{figure}

To reproduce and investigate the circular band formation similar to those 
observed in experiments \cite{Lipson2002, Seiden2005}
for the suspensions of particles in a fully filled horizontally rotating cylinder, 
we have considered the cases of 16, 24, 32, and 64 balls of radius $a=$0.075 cm and density 
$\rho_p =$ 1.25 g/cm$^3$ in a truncated cylinder of diameter $2R=1$ cm and length 4 cm filled 
with a fluid of the density 1 g/cm$^3$ and kinetic viscosity $\nu=0.15$ cm$^2$/sec.
The initial positions of the ball mass centers are on the circles of radius 0.35 cm centered 
at the central axis of the cylinder with eight balls in each circle (see Figures \ref{fig.3}-\ref{fig.6}).
We have perturbed each mass center randomly in the direction of the cylinder axis
to break the symmetry of the initial pattern. The distance between two neighboring circles is 
about 2.25$a$ hence the initial gap size $d_g$ between balls in the cylinder axis direction is 
about $a/4$. In the simulations, the cylinder rotates about the cylinder axis parallel to the 
$y$-axis  in a clockwise direction with rotating rate $\Omega$ equal to either 8 or 12 
sec$^{-1}$ (see Figure  \ref{fig.1}).  Then the  associated values of Reynolds number, 
$Re_p=2a R \Omega /\nu$, are 4 and 6, respectively.

\begin{figure}
\begin{center}
\leavevmode
\epsfxsize=2.7in
\epsffile{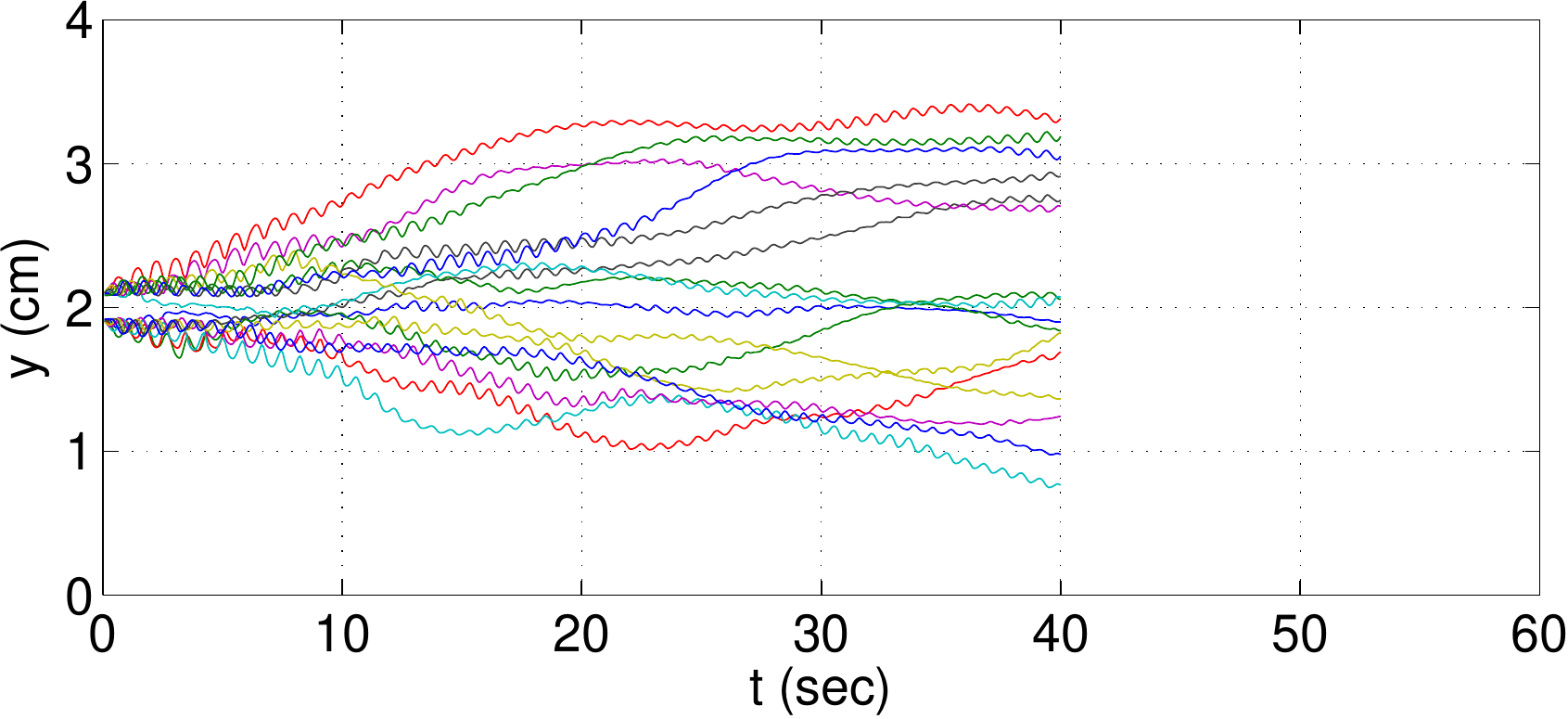}
\epsfxsize=2.7in
\epsffile{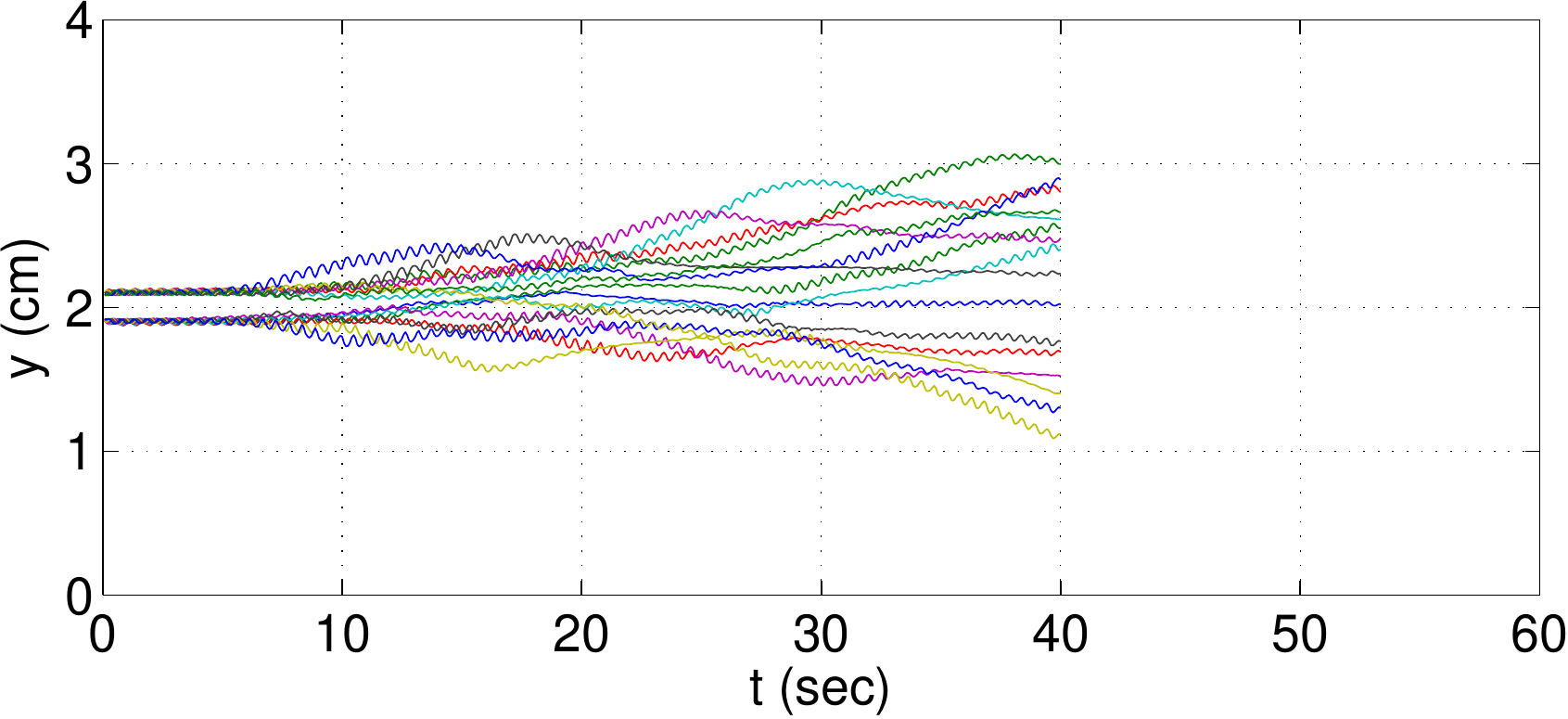}\\
\epsfxsize=2.7in
\epsffile{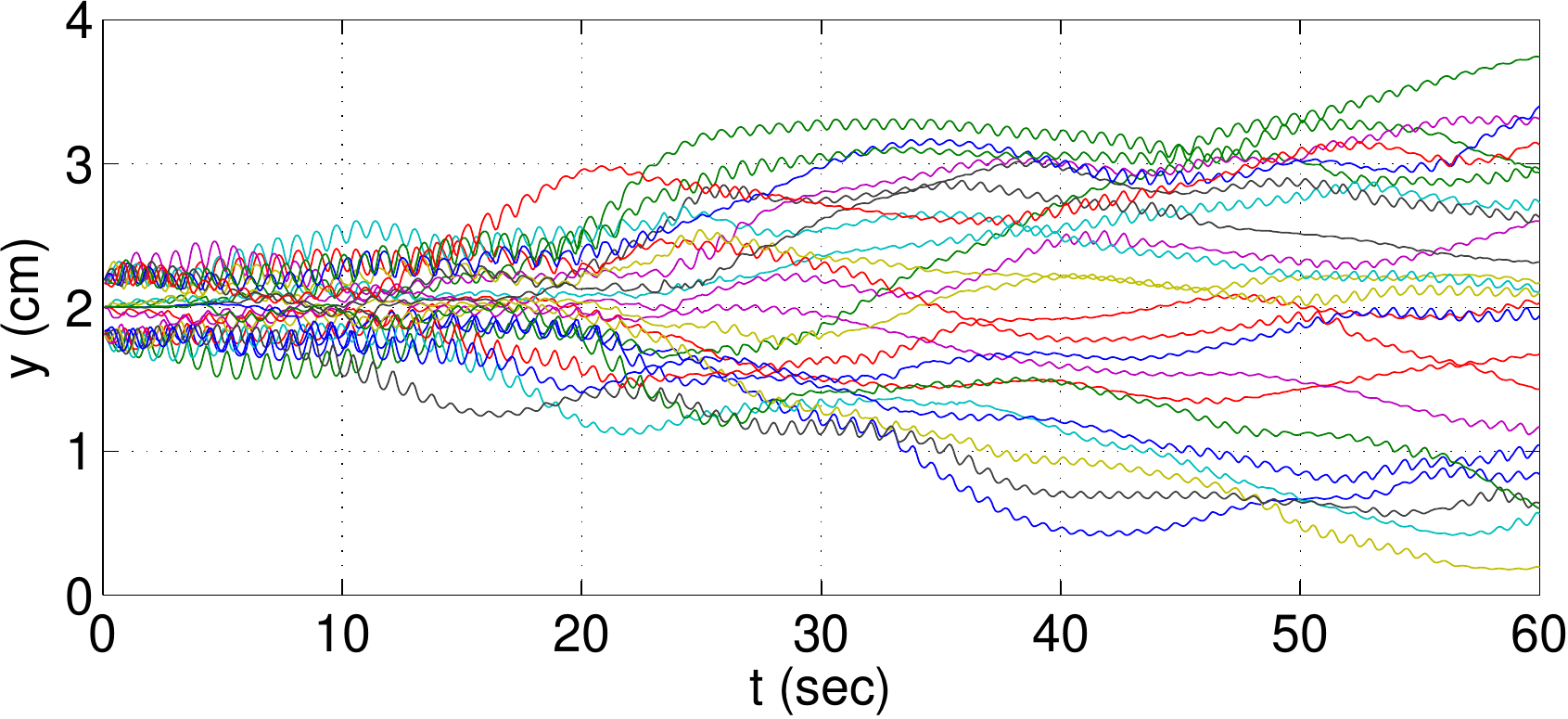}
\epsfxsize=2.7in
\epsffile{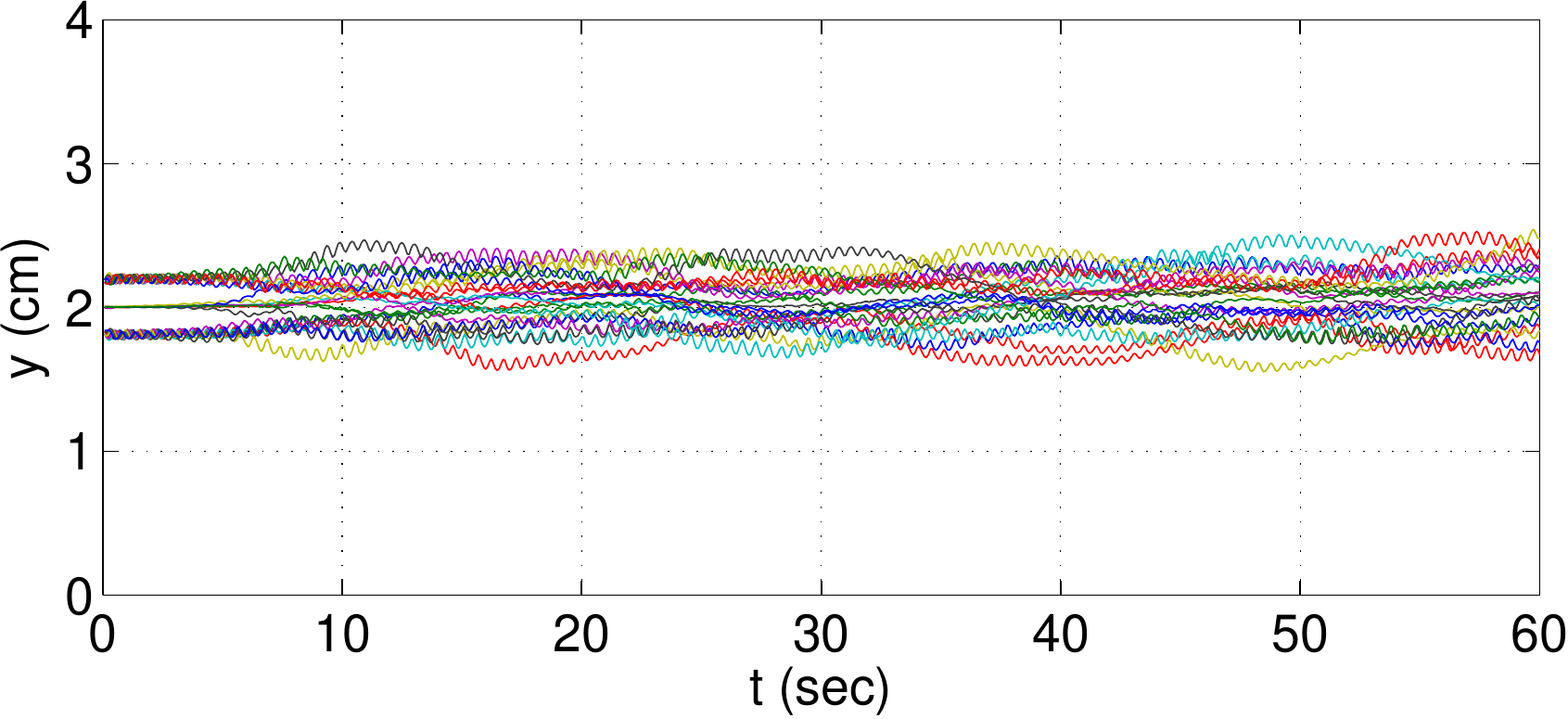}\\
\epsfxsize=2.7in
\epsffile{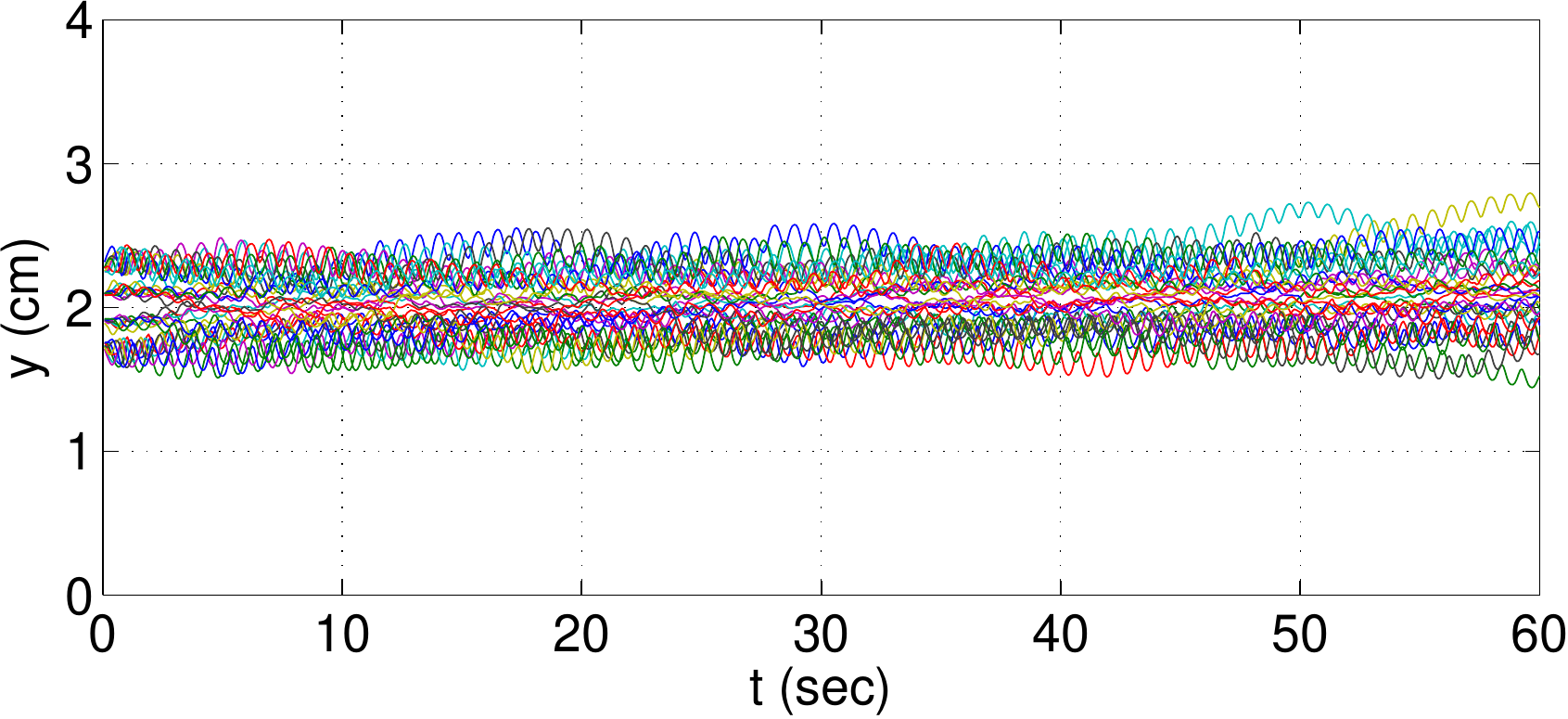}
\epsfxsize=2.7in
\epsffile{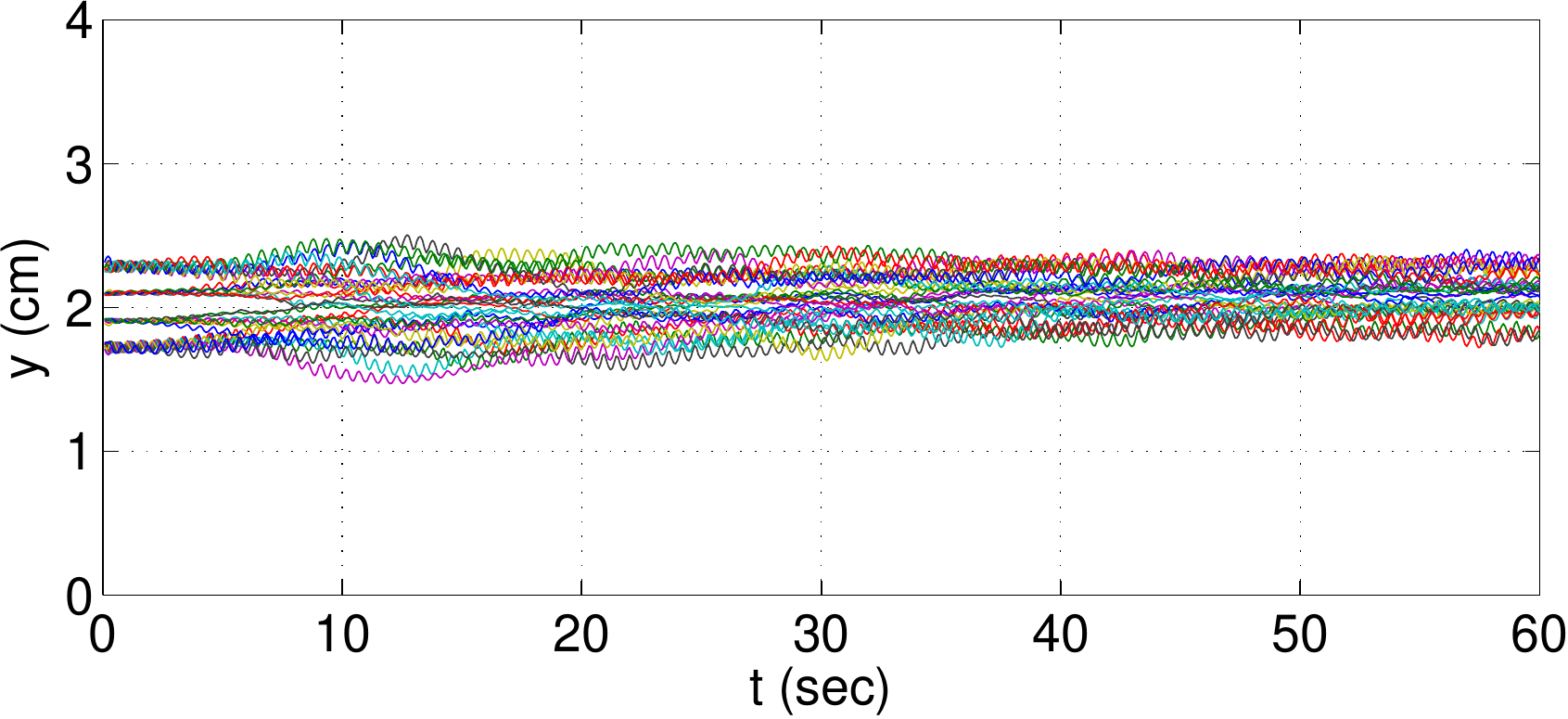}\\
\epsfxsize=2.7in
\epsffile{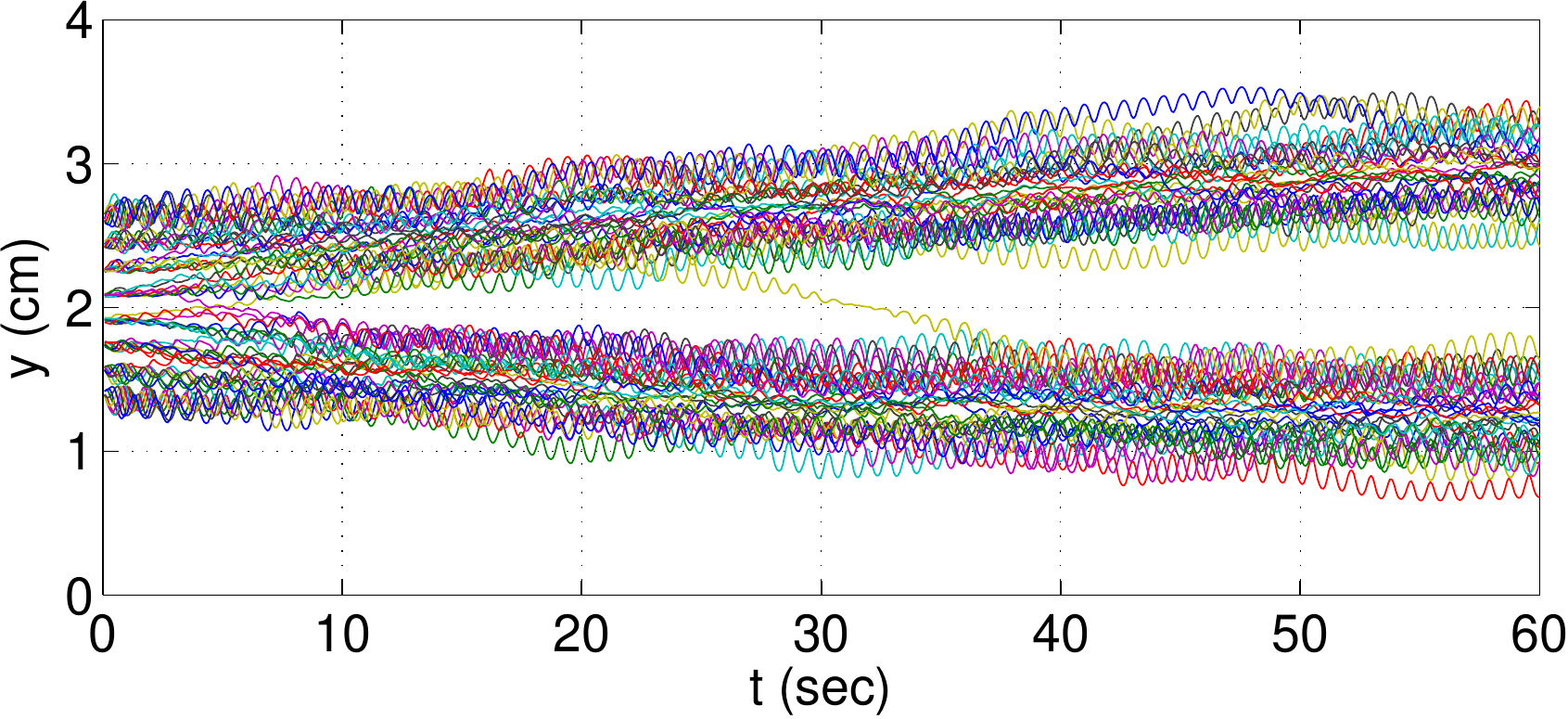}
\epsfxsize=2.7in
\epsffile{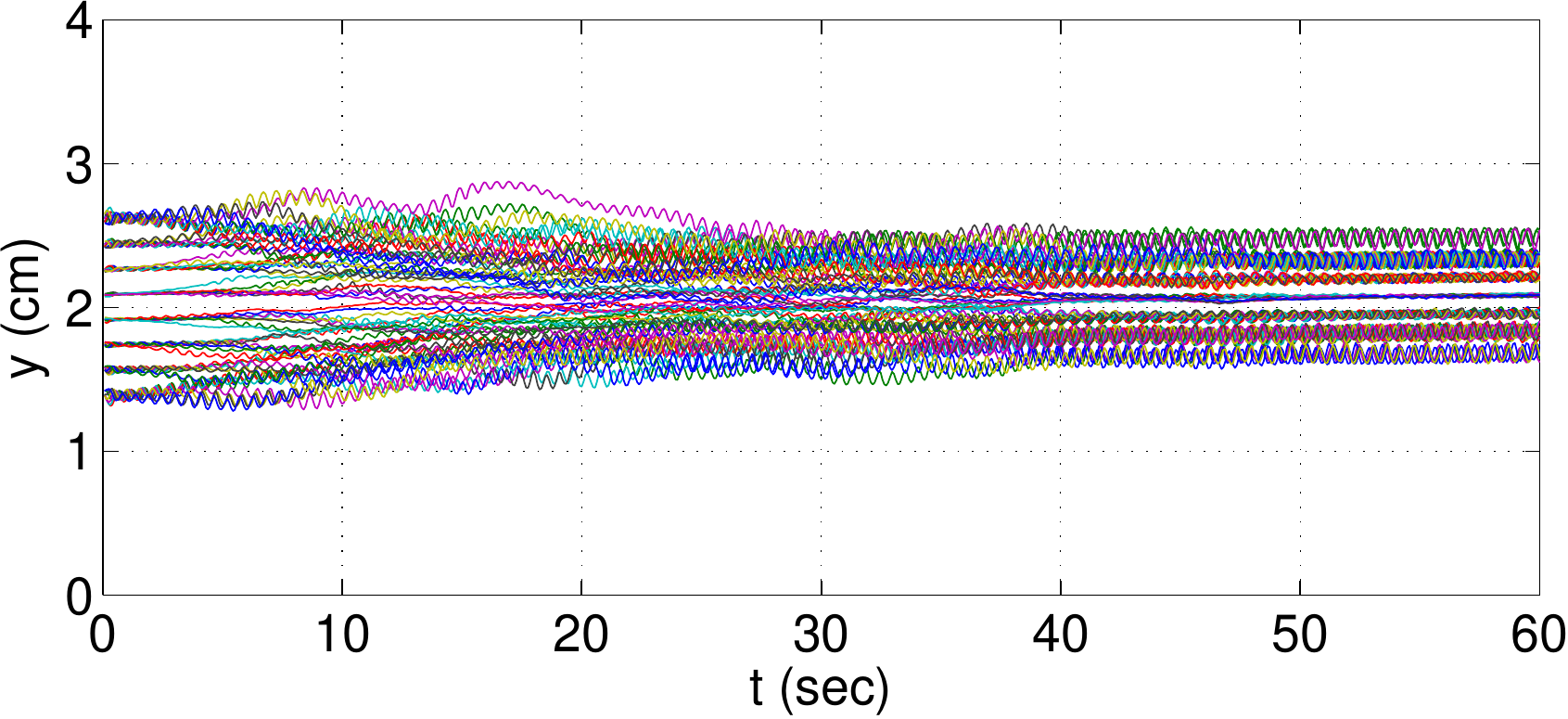}
\end{center}
\caption{The histories of the $y$-coordinates of the mass centers of 16, 24, 32, and 64 balls
(from top to bottom): $\Omega=$ 8 (left) and 12 (right) sec$^{-1}$.}\label{fig.7}
\begin{center}
\leavevmode
\epsfxsize=2.25in
\epsffile{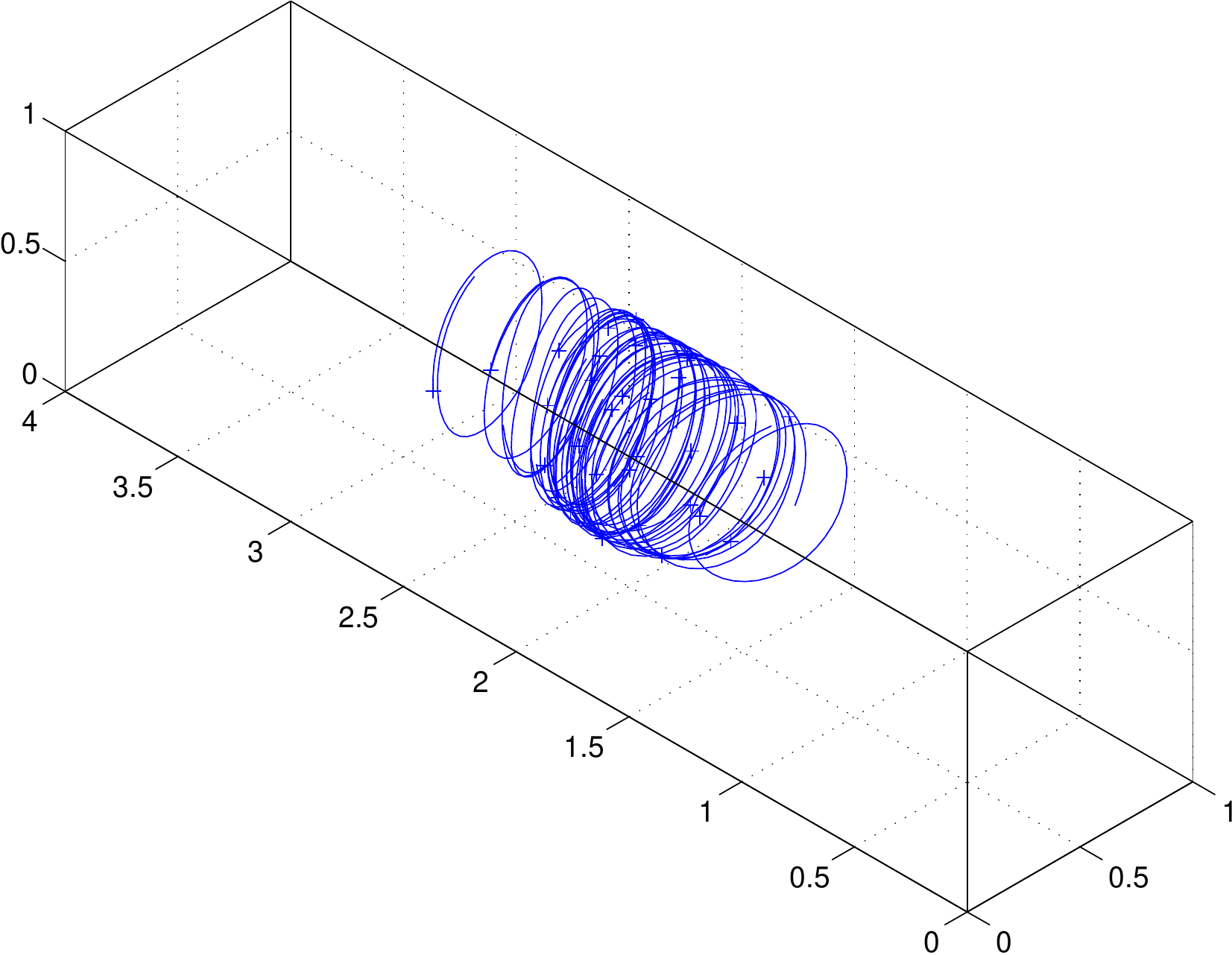}
\epsfxsize=0.65in
\epsffile{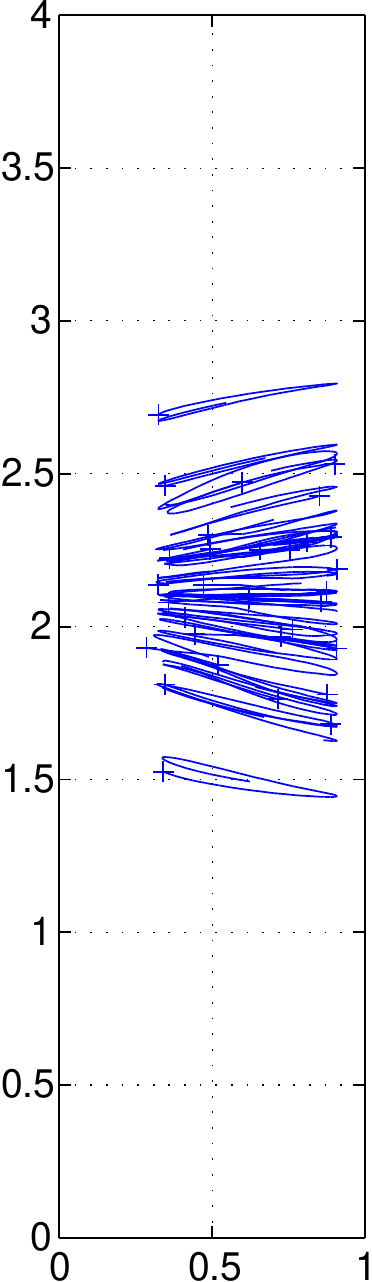}
\epsfxsize=2.25in
\epsffile{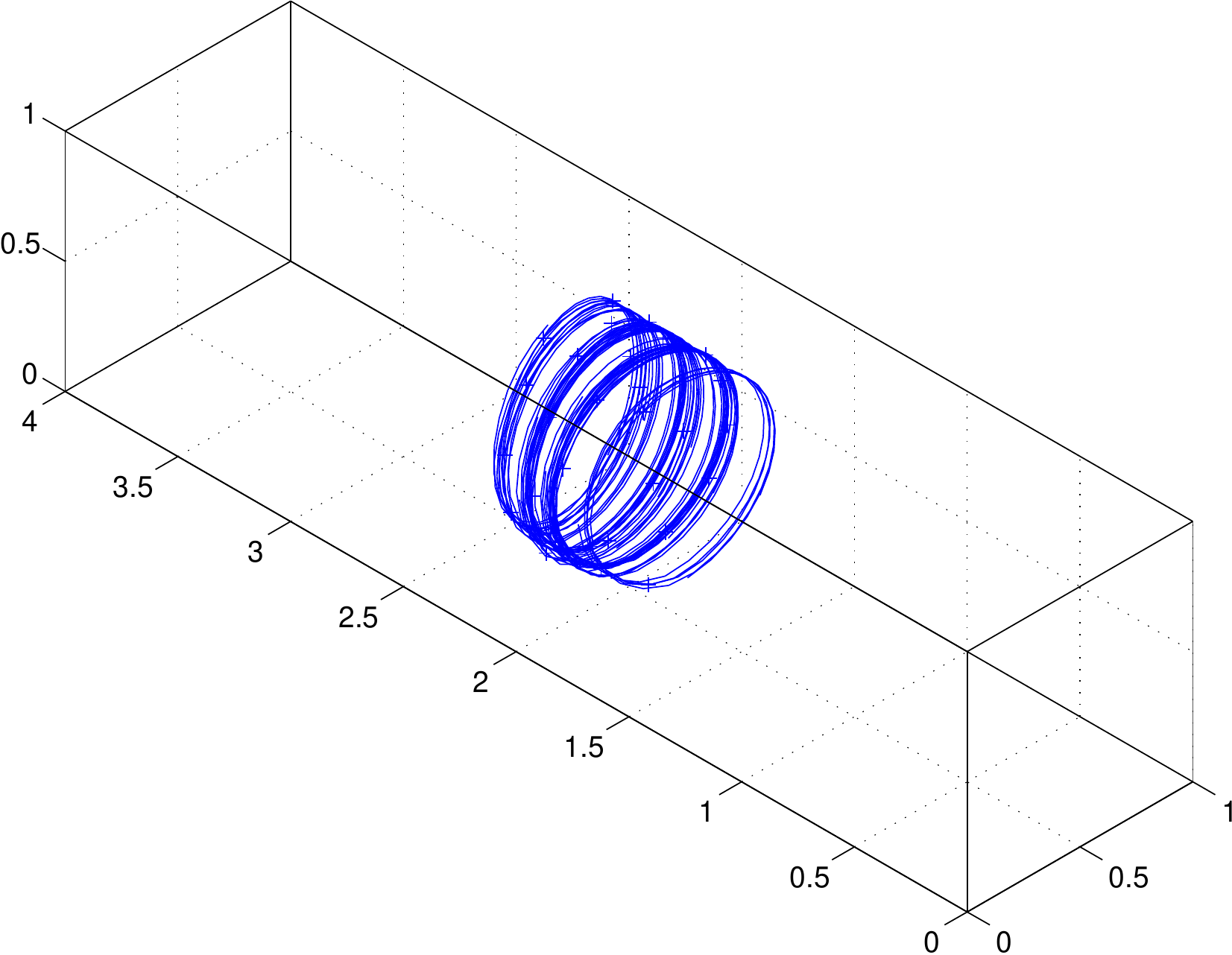}
\epsfxsize=0.65in
\epsffile{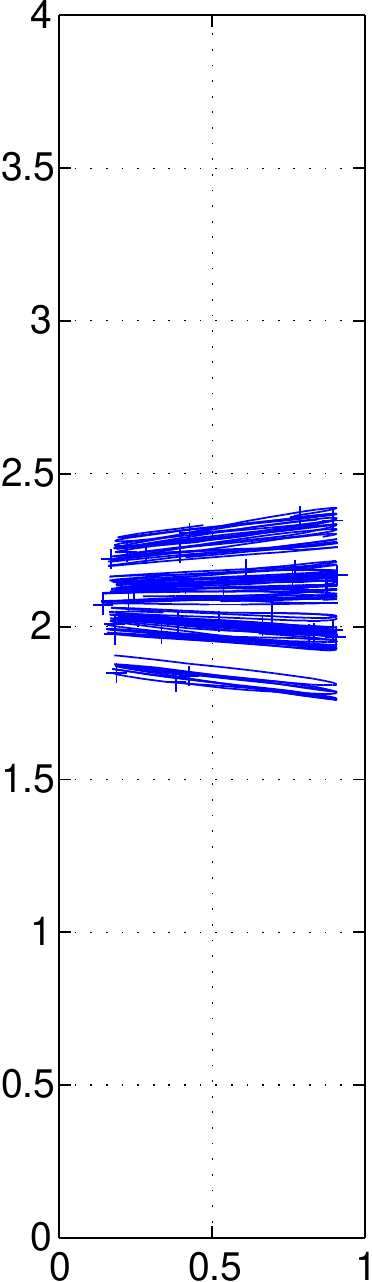}
\end{center}
\caption{The trajectories and the top view of the trajectories of the 32 balls at the rotating rate 
$\Omega=$ 8 (left two) and 12 (right two) sec$^{-1}$ for $59 \le t \le 60$.} \label{fig.8}
\end{figure}

The  histories of the $y$-coordinates of the particle  mass centers  and the positions of 16 balls 
at $t=40$ second obtained with the rotating rates of $\Omega =$ 8 and 12 sec$^{-1}$ in Figures \ref{fig.3} 
and \ref{fig.7} clearly show  that the 16 balls spread out in the cylinder axis direction and 
do not form a compact circular band at all. For the cases of 24 balls, the formation of the 
circular band is still not clear yet. In Figures \ref{fig.4} and \ref{fig.7}, the 24 balls 
spread out in the cylinder axis direction at the rotating rate $\Omega =$ 8 sec$^{-1}$. When 
the rotating rate is 12 sec$^{-1}$, the 24 balls do form a loosely circular band.  For the cases 
of 32 balls, the formation of the circular band is clearly shown in Figures \ref{fig.5} and \ref{fig.7}. 
The one obtained at the rotating rate $\Omega =$12 sec$^{-1}$ is very compact. When having 64 balls, 
they split into two loosely circular bands at $\Omega =$ 8  sec$^{-1}$ since the rotating rate is not 
fast enough to produce strong particle interaction to sustain the whole group of particles as shown
in Figures \ref{fig.6} and \ref{fig.7}. But at the rotating rate $\Omega =$ 12 sec$^{-1}$, 
there is just one compact circular band and the particles are well organized in the middle of the 
group by the pushing from the outer particles. The particles form a layer inside the cylinder 
which is different from those observed in \cite{Matson2003, Matson2005, Matson2008, Breu2003}, 
but close to those in \cite{Lipson2002,Seiden2005} . 
These results give us a simple observation which is that there is a need of enough particles so 
that the particles within a circular band can continuously interact among themselves. For the cases 
of 64 particles shown in Figures \ref{fig.6} and \ref{fig.11}, there are 33 and 31 particles in two bands, respectively, in Figure 6 and 29 and 35 in two bands, respectively, in Figure 11. We believe that
about 30 particles are enough to form a circular band for the conditions considered in this paper.


By observing the trajectories of 32 balls  in Figure \ref{fig.8}, 
we have found that the ball trajectories become closer to each other when the balls move 
downward at the back of the cylinder ($x=0$) and they loose up (spreading out in the
cylinder axis direction) when the balls move upward at the front of the cylinder ($x=1$).
A typical hydrodynamical interaction between balls called the drafting, kissing and 
tumbling phenomenon (e.g., see \cite{Fortes1987}), due to the fluid flow inertia, 
plays a role here in a fully filled horizontally rotating cylinder. When a circular band 
is forming, the part of a circular band formed by the balls moving downward becomes more 
compact due to the effects of the speedup caused by the gravity and the hydrodynamical 
interaction between balls. For the  part of a circular band formed by the balls moving 
upward, it is loosening up due to the slow down of the particle motion by the counter 
effect of the gravity and its width becomes wider (see Figure \ref{fig.8}). Also due to 
these effects of the speedup and slow down, the histories of the  $y$-coordinates of the 
particle mass centers  in Figure \ref{fig.7} show oscillations. To have a stabilized and 
compact circular band, enough number of balls and a fast rotating rate are needed in order to
balance both effects, e.g., the results of the   32 ball cases  in Figures \ref{fig.5} and 
\ref{fig.8} show that at the rotating rate $\Omega =$ 8 sec$^{-1}$, the particle speeds are not 
fast enough to overcome the diffusion caused by the gravity when they move upward at the front 
of the cylinder. But at the rotating rate $\Omega =$ 12 sec$^{-1}$, the particle speeds are 
fast enough so that a compact circular band is formed. Similarly for the 64 ball case at
the rotating rate $\Omega =$ 8 sec$^{-1}$ in Figures \ref{fig.6} and \ref{fig.7}, the particle
interaction can only hold particles in two loosely circular bands; but at  $\Omega =$ 12 sec$^{-1}$ 
the particle interaction can pull all 64 particles together in one compact circular band.



\subsection{The effect of the band formation on the fluid flow field}

\begin{figure}
\begin{center}
\leavevmode
\epsfxsize=2.9in
\epsffile{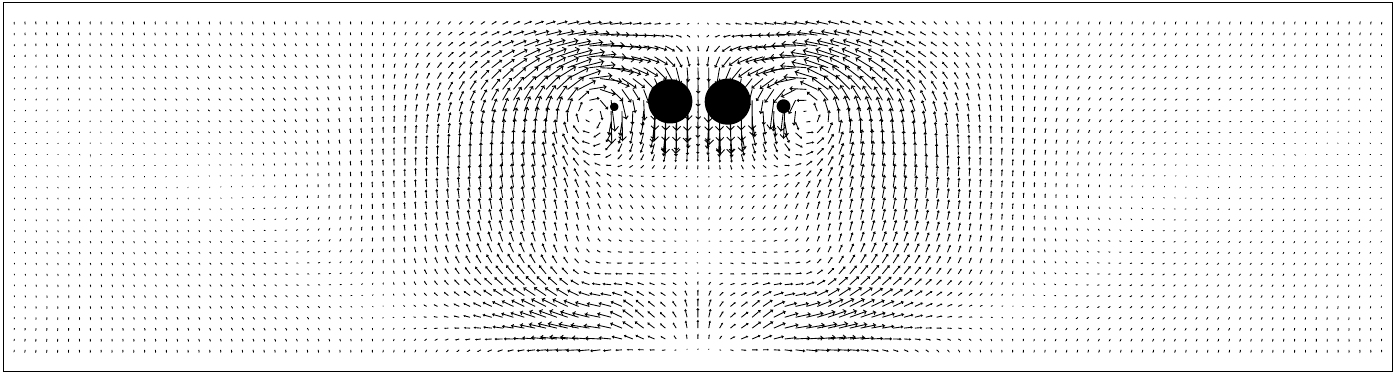} \hskip 10pt
\epsfxsize=3.in
\epsffile{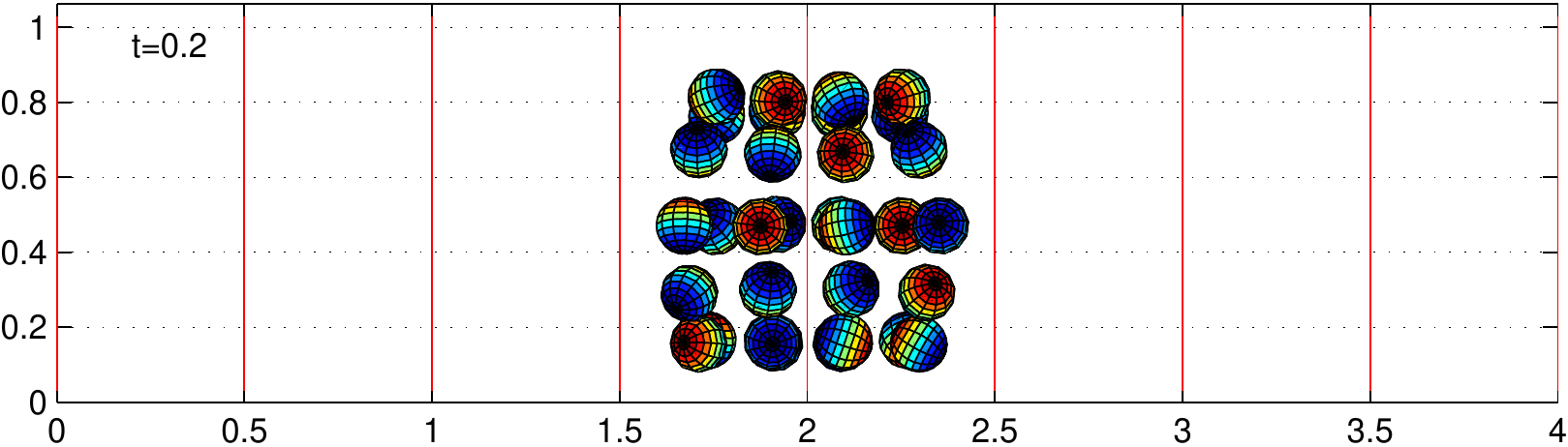}\\
\epsfxsize=2.9in
\epsffile{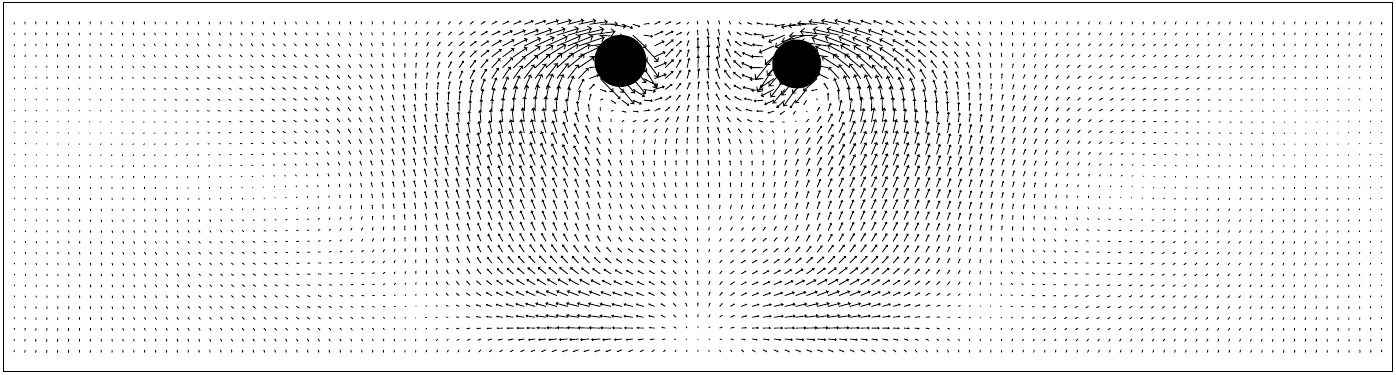} \hskip 10pt
\epsfxsize=3.in
\epsffile{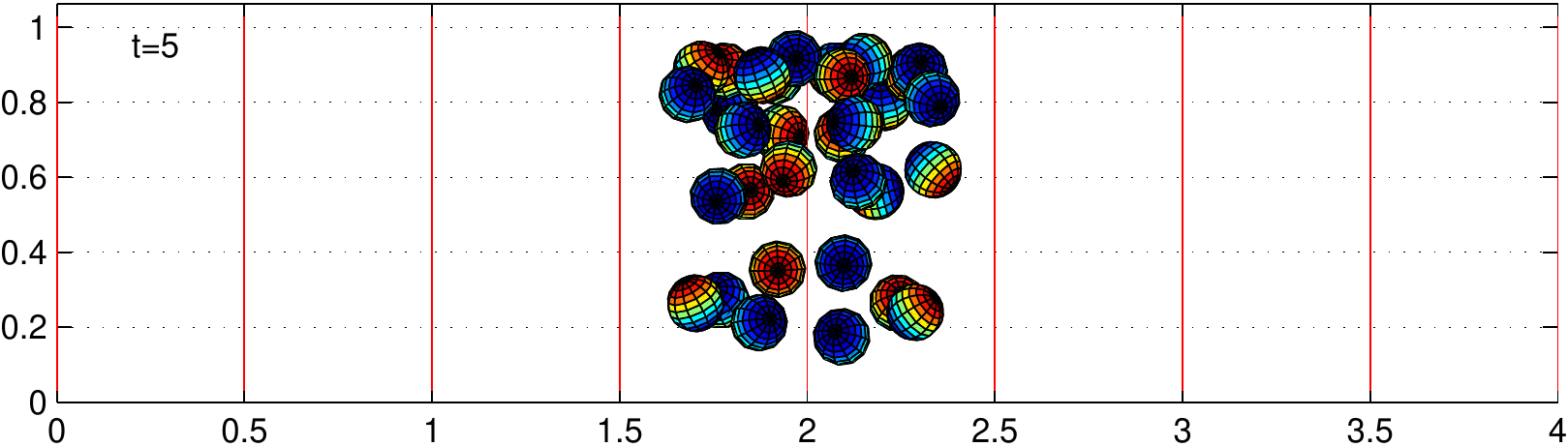}\\
\epsfxsize=2.9in
\epsffile{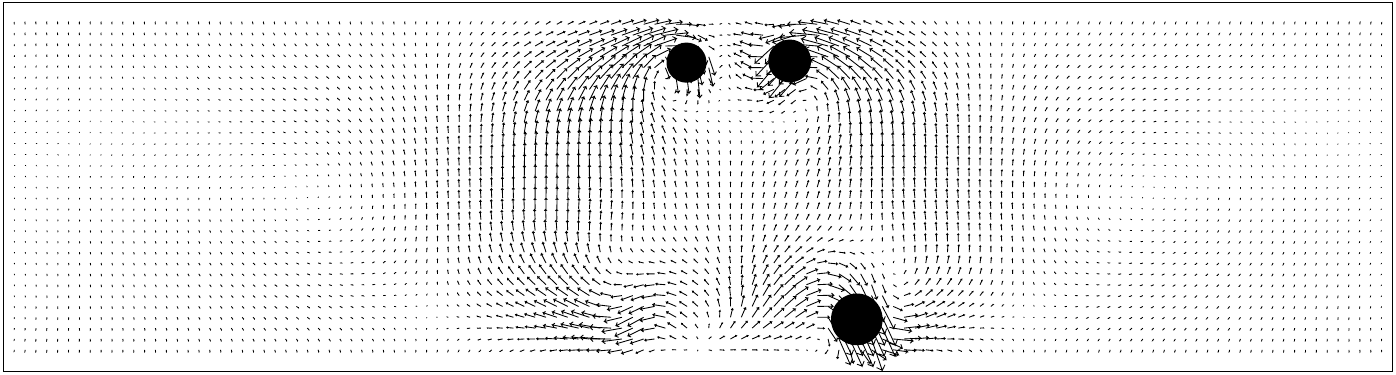} \hskip 10pt
\epsfxsize=3.in
\epsffile{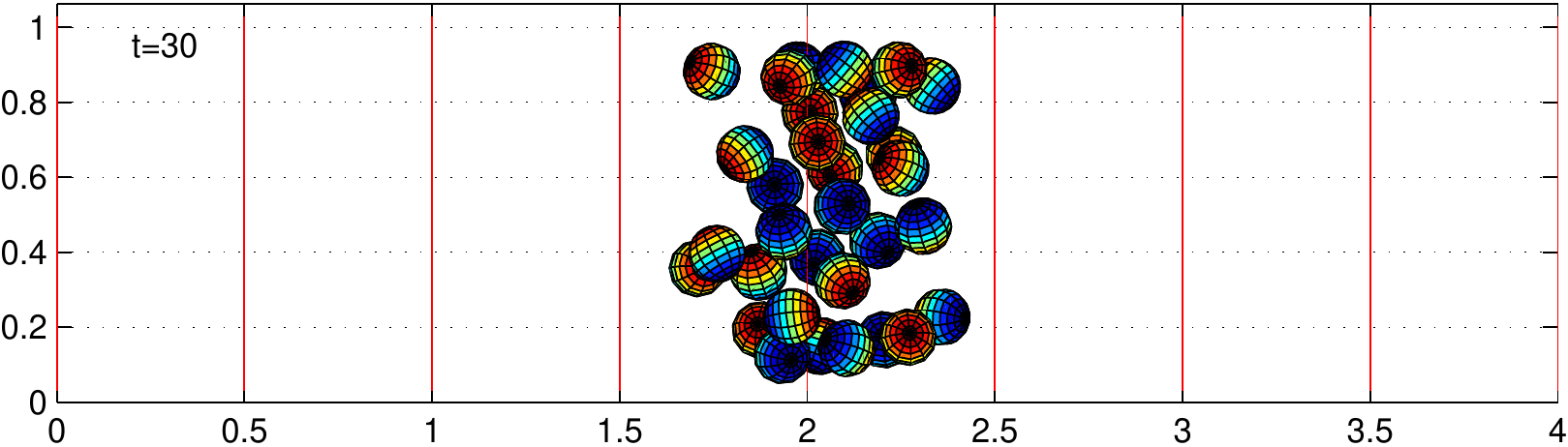} \\
\end{center}
\caption{(a) The projection of the velocity field on the vertical plane passing through the central
axis of the cylinder for the  case of 32 balls (left) and (b) the front view  of the position of 32 balls (right) at $t=$ 0.2, 5  and 30 second (from top to bottom) with $\Omega=$ 12 sec$^{-1}$.}\label{fig.10}
\begin{center}
\leavevmode
\leavevmode
\epsfxsize=2.9in
\epsffile{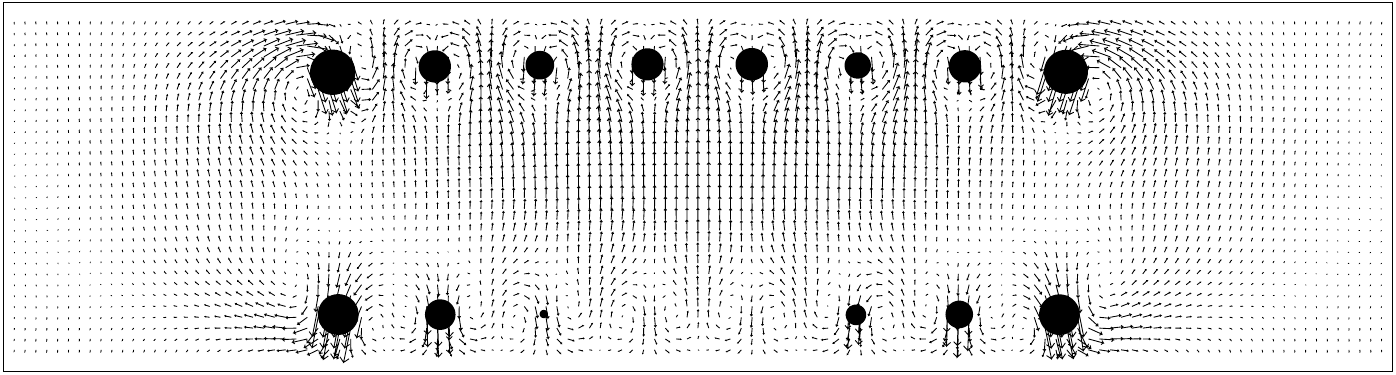}  \hskip 10pt
\epsfxsize=3.in
\epsffile{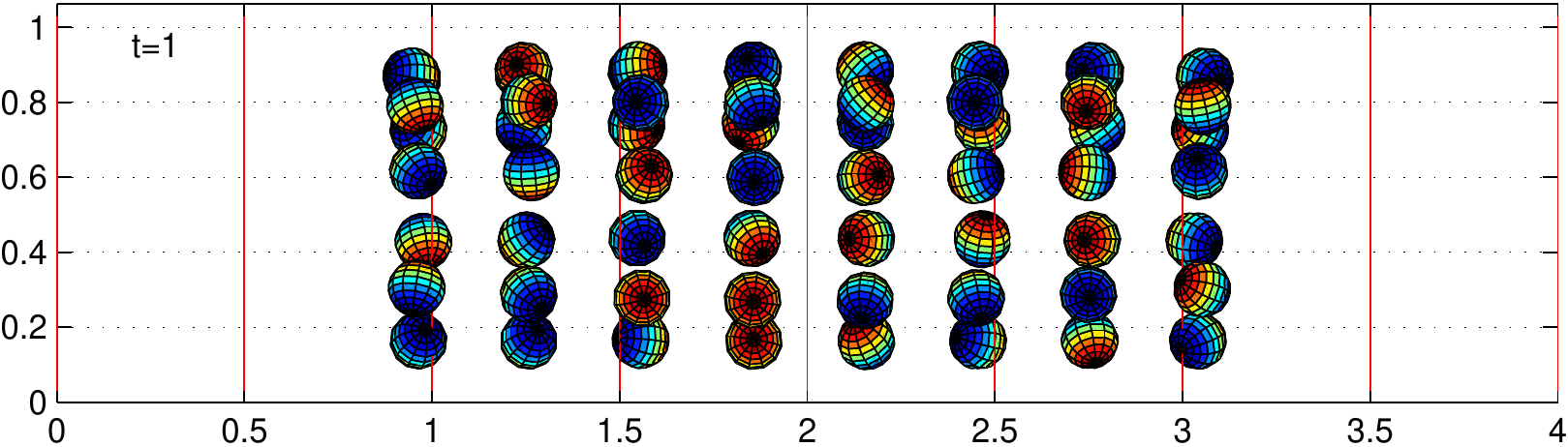}\\
\epsfxsize=2.9in
\epsffile{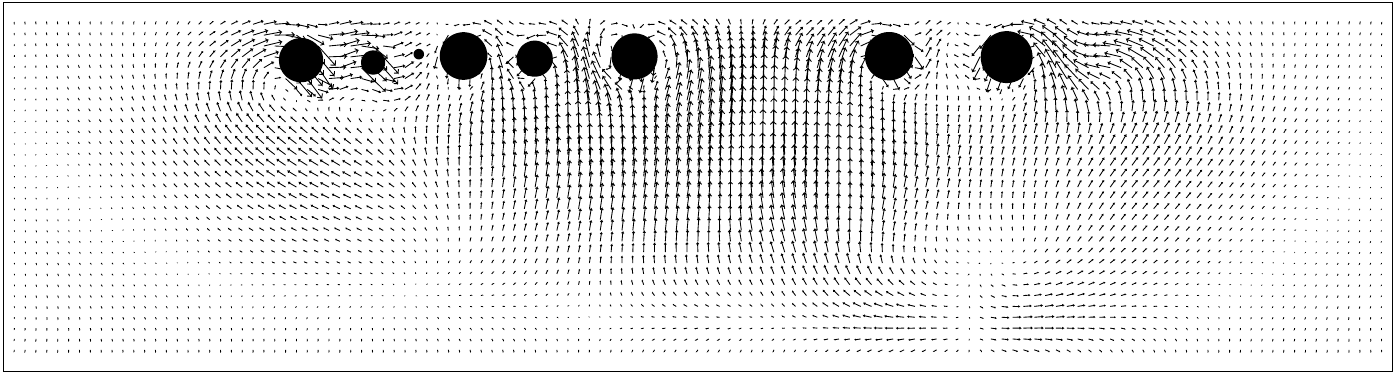}  \hskip 10pt
\epsfxsize=3.in
\epsffile{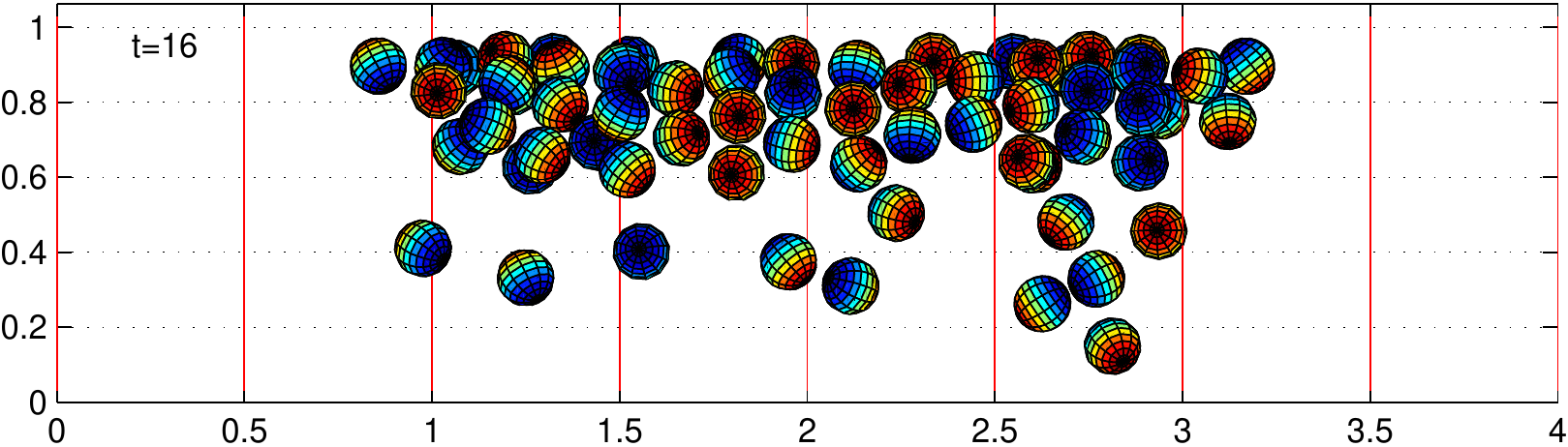}\\
\epsfxsize=2.9in
\epsffile{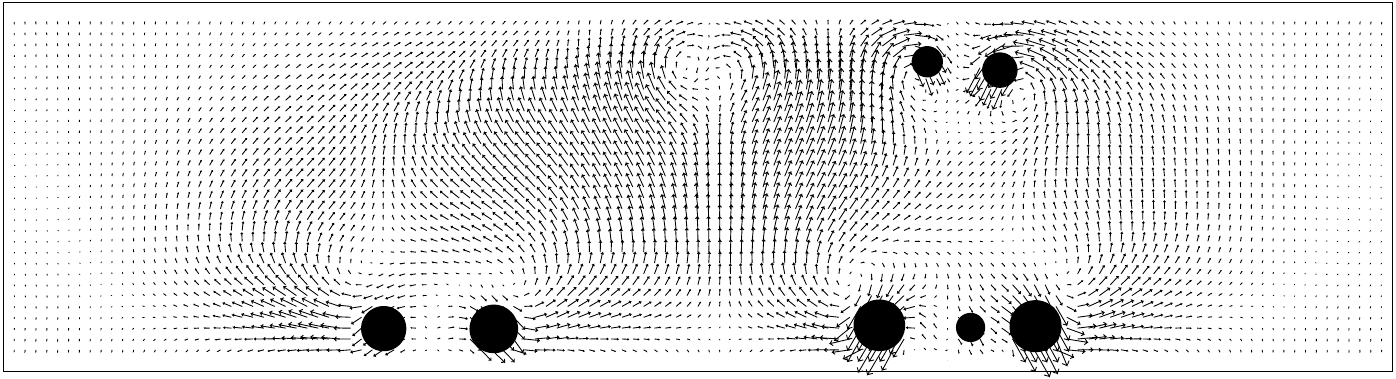}  \hskip 10pt
\epsfxsize=3.in
\epsffile{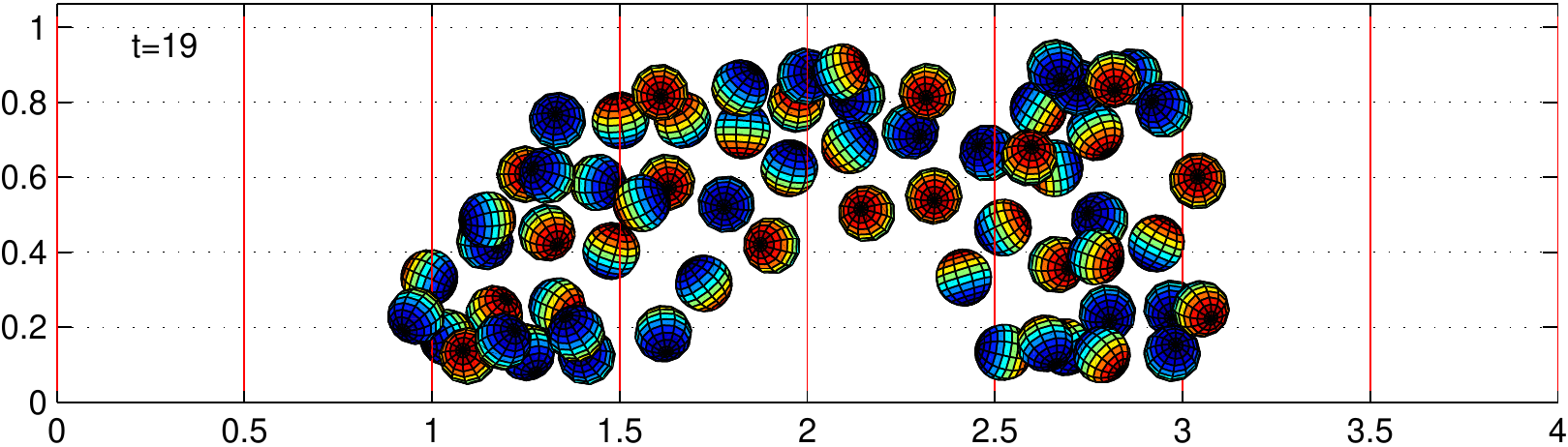}\\
\epsfxsize=2.9in
\epsffile{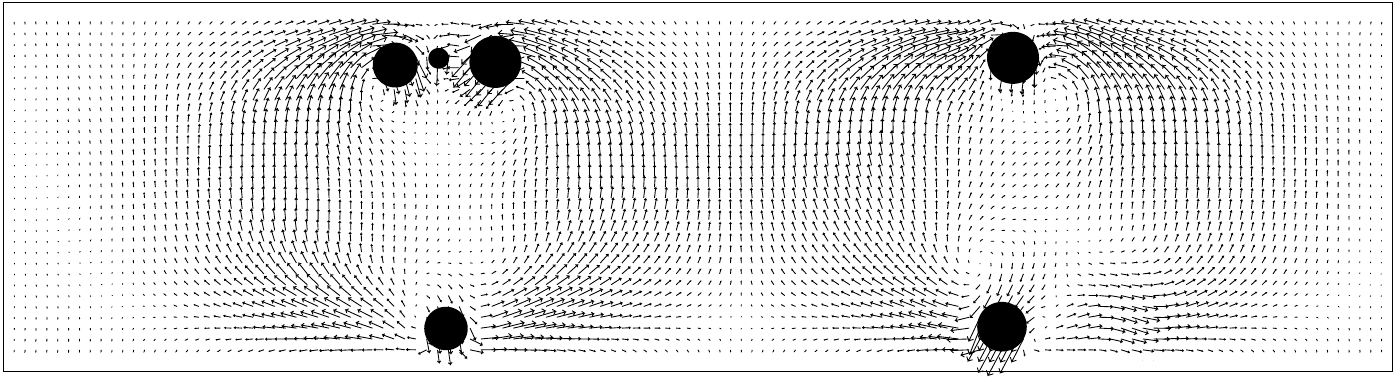}  \hskip 10pt
\epsfxsize=3.in
\epsffile{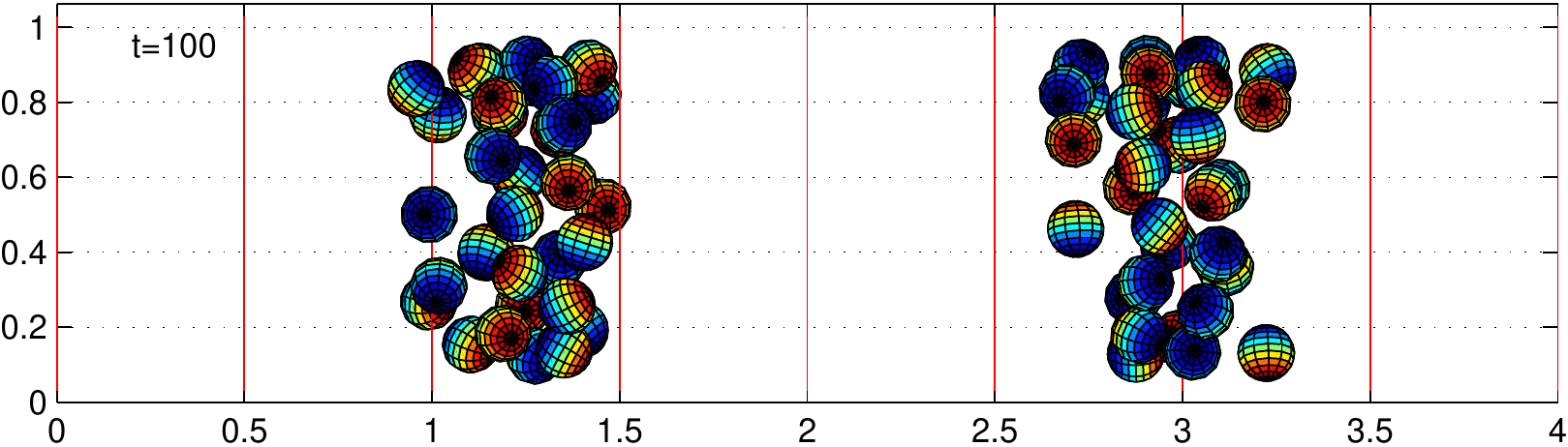}
\end{center}
\caption{(a) The projection of the velocity field on the vertical plane passing through the central
axis of the cylinder for the  case of 64 balls (left) and (b) the front view  of the position of 64 balls  (right) at $t=$ 1, 16, 19 and 100 second (from top to bottom) with $\Omega=$ 12 sec$^{-1}$ and 
the initial gap size $d_g=2 a$.}\label{fig.11}
\end{figure}

\begin{figure}[ht]
\begin{center}
\leavevmode
\leavevmode
\epsfxsize=6in
\epsffile{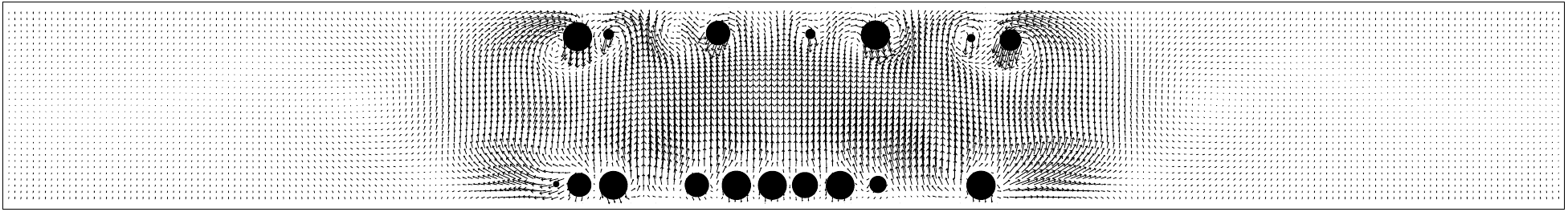}\\
\epsfxsize=6in
\epsffile{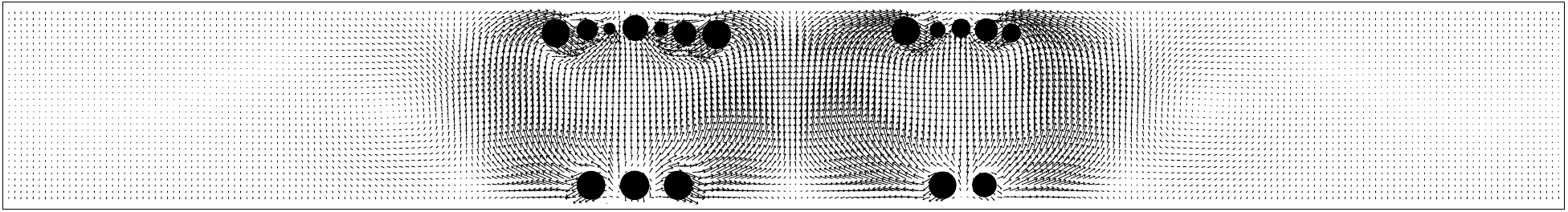}\\
\epsfxsize=6.in
\epsffile{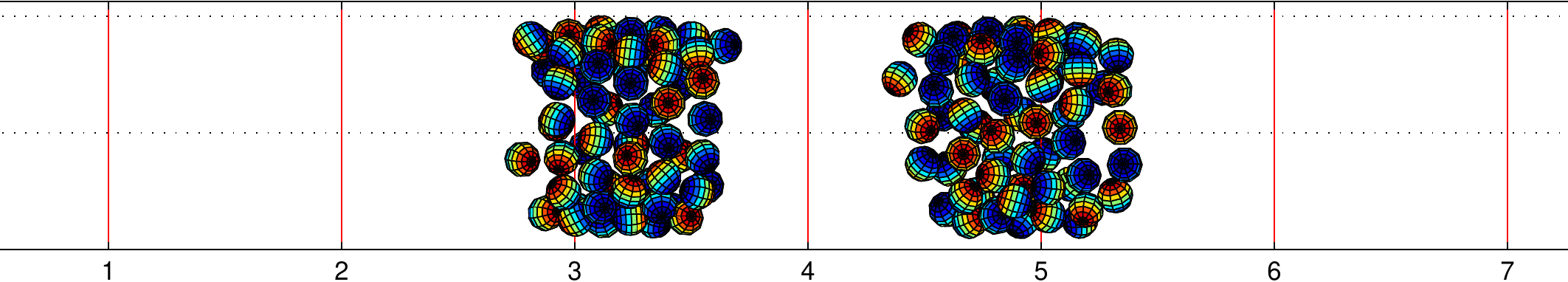} 
\end{center}
\caption{The projection of the velocity field on the vertical plane passing through the central
axis of the cylinder for the  case of 128 balls at  $t=$ 6 (top) and 40 second (middle) 
and the front view  of the position of  128 balls  (bottom) at $t=$ 40 second.}\label{fig.12}
\begin{center}
\leavevmode
\epsfxsize=3.0in
\epsffile{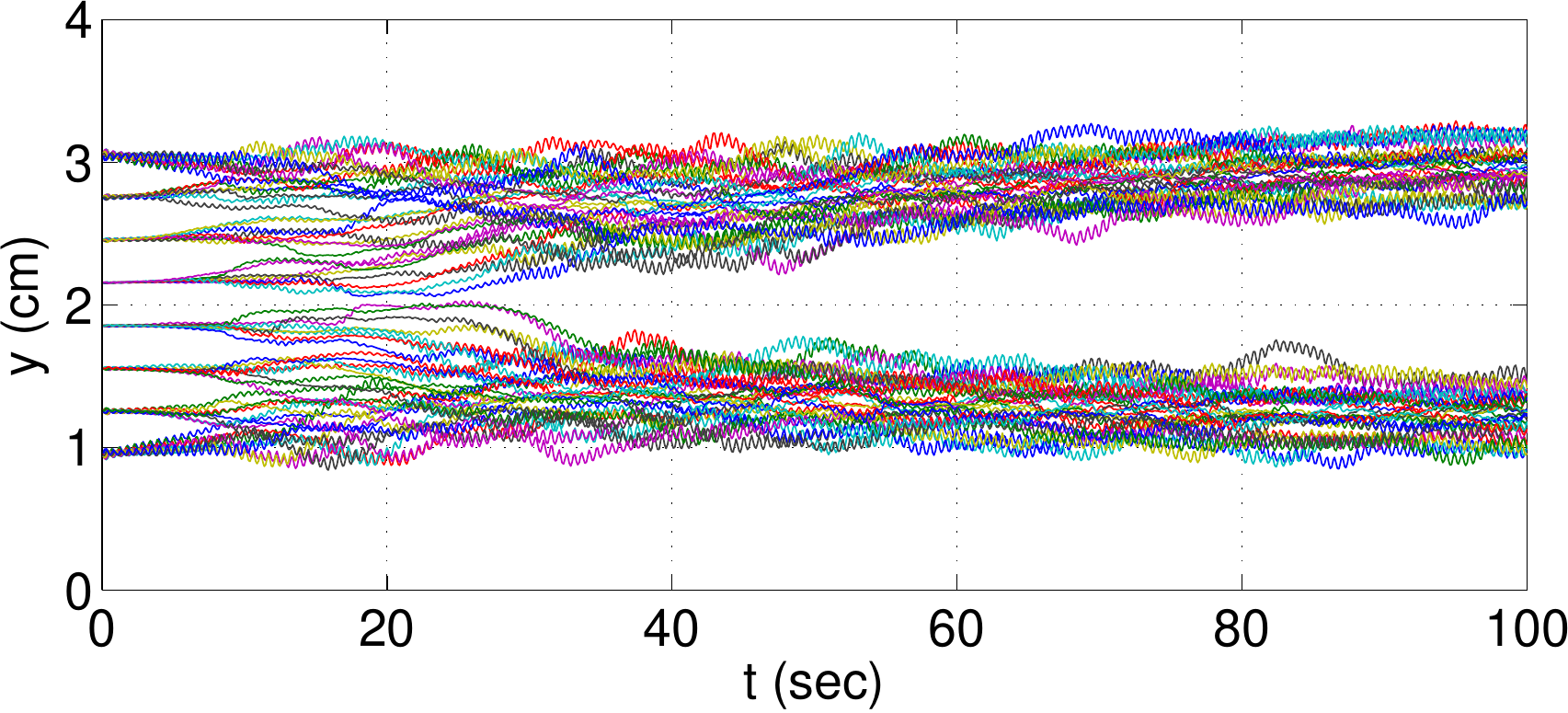}
\epsfxsize=3.0in
\epsffile{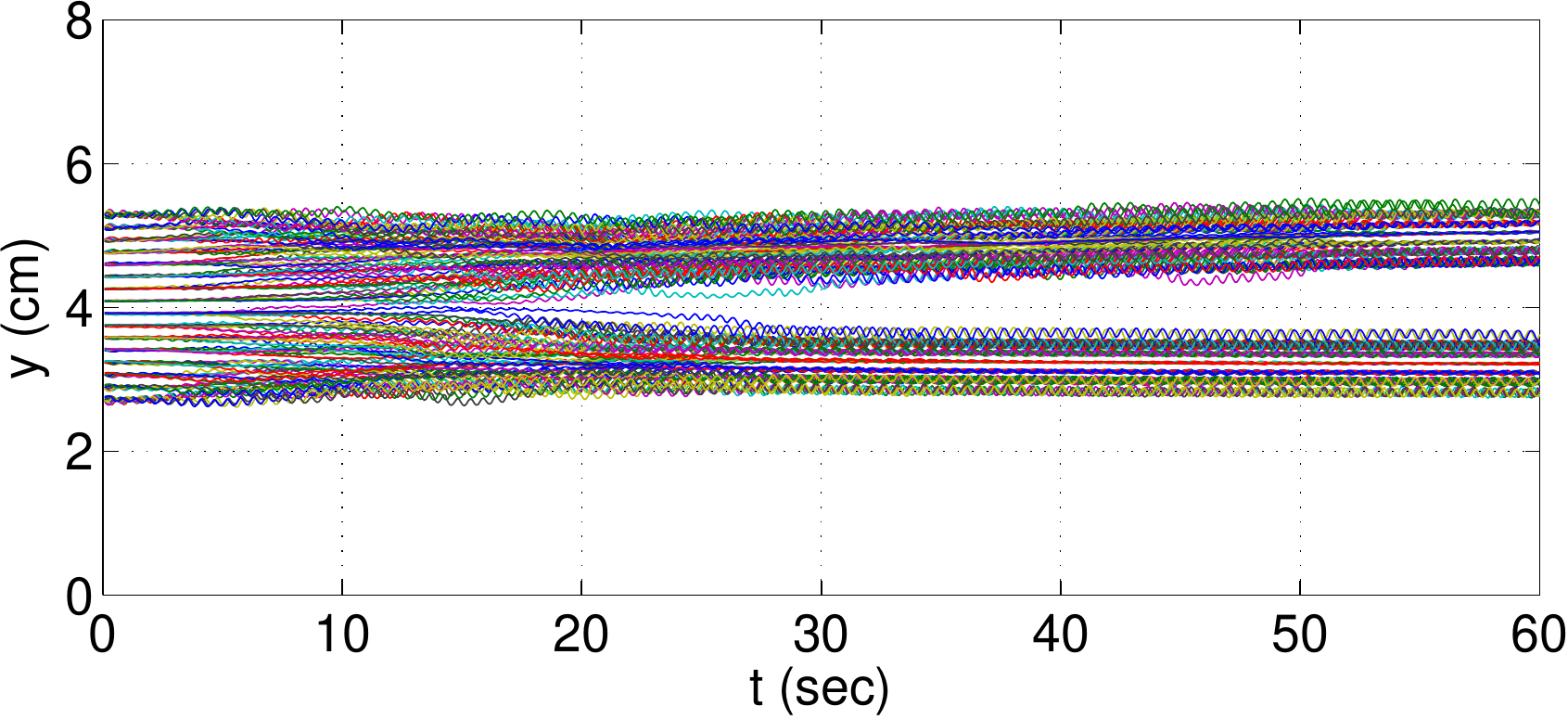}
\end{center}
\caption{The histories of the $y$-coordinates of the mass centers of 64 balls with
the initial gap size $d_g=2 a$ (left) and 128 balls with
the initial gap size $d_g=a/4$ (right).}\label{fig.13}
\end{figure}

In \cite{Seiden2005}, Seiden et al. suggested that the formation of circular 
bands (e.g., in Figure 13 in \cite{Seiden2005}) is resulted by  mutual interaction between the 
particles and the periodic inertial waves in the cylinder axis direction. 
Via investigating the flow field associated from the results obtained in the previous subsection, 
we have different point of view concerning the formation of circular bands involving 
the periodic inertial waves  in the cylinder axis direction. The initial positions of 
the particles in numerical simulations are not easy to set up experimentally; but such 
kind of arrangement does help us to understand the formation of circular bands. 
For the case of 32 balls and $\Omega=12$ sec$^{-1}$ studied in the previous subsection, 
the projections of the velocity field on the vertical plane passing through the central 
axis of the cylinder at different time are shown in Figure \ref{fig.10}. The 32 particles 
do not excite the entire flow field inside the rotating cylinder into a periodic flow 
field in the cylinder axis direction at all.  The circulation of the velocity field is 
created only by the particle motion and concentrated in the middle portion of the cylinder. 
For the evolution of the flow field related to two circular bands, the results of the case 
of 64 balls with the initial gap size $d_g=2a$ and the rotating rate $\Omega =$ 12 sec$^{-1}$ 
are shown in Figures \ref{fig.11} and \ref{fig.13}.  We have observed no periodic flow field 
from the beginning as in Figure \ref{fig.11}. The particles  break into two circular bands 
between $t =$ 16 and 19 second and then two bands move away from each other as in Figure \ref{fig.13} 
and the projected velocity fields at $t=$ 19 and 100 second in Figure \ref{fig.11} show that 
the two circulations move apart since the two circular bands move away from each other. 
The projected velocity field at $t=$ 100 second in Figure \ref{fig.11} is similar to the one 
in Figure 13 in \cite{Seiden2005} obtained experimentally in \cite{Seiden2005}, but the circulation of the 
flow field is caused by the motion of the particles in the two circular particle bands. Similarly 
for the case of 128 balls in a truncated cylinder of length $L=$8 cm at the rotating rate 
$\Omega=12$ sec$^{-1}$ with the initial gap size $d_g=a/4$, the particles are initially placed 
on 16 circles in the middle of the cylinder as in the previous subsection. Later they break into two 
compact circular bands as shown in Figures \ref{fig.12} and \ref{fig.13}. There are 63 and 65 
particles in these two circular bands, respectively, which are consistent with the results of 
the 64 particles at the rotating rate $\Omega=12$ sec$^{-1}$ discussed in the previous subsection. 
The figures of the circulation of the flow field  at $t=$ 6 and 40 second  in Figure \ref{fig.12} 
clearly show that there is no periodic flow field pattern in the cylinder axis direction, but 
there is only one large circulation at the beginning and then two circulations are
created by the two particle bands once the particles break into two groups.


\begin{figure}[ht]
\begin{center}
\leavevmode
\hskip 3.1in
\epsfxsize=3.in
\epsffile{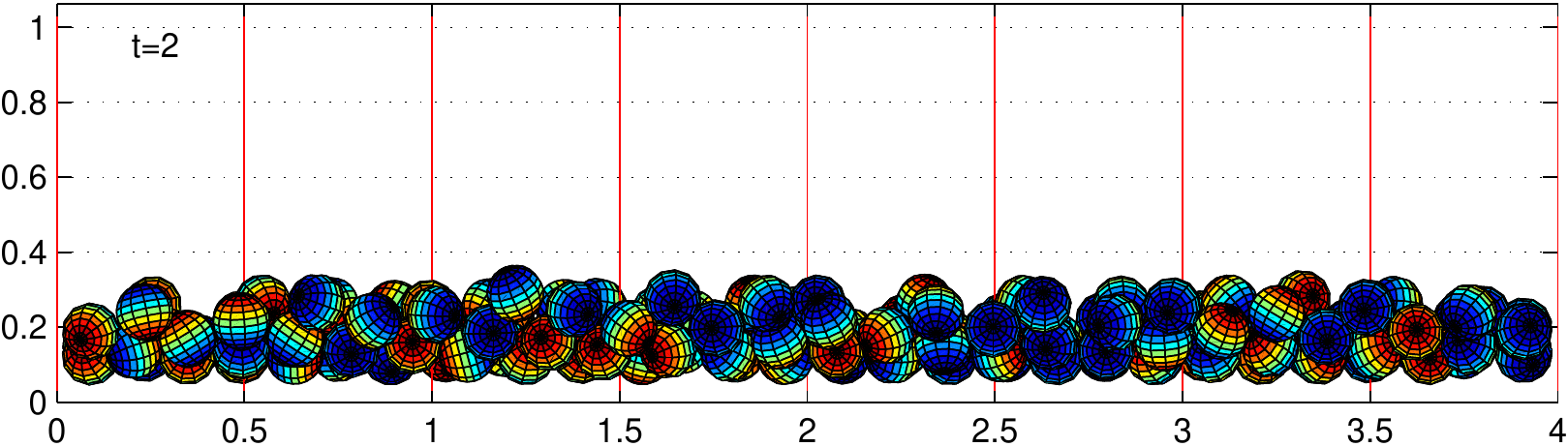}\\
\epsfxsize=2.9in
\epsffile{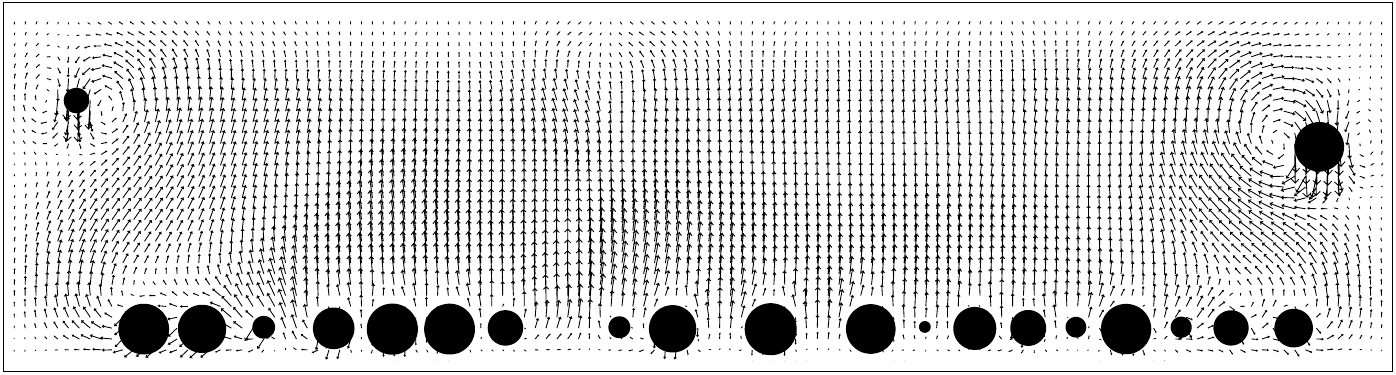}  \hskip 10pt
\epsfxsize=3.in
\epsffile{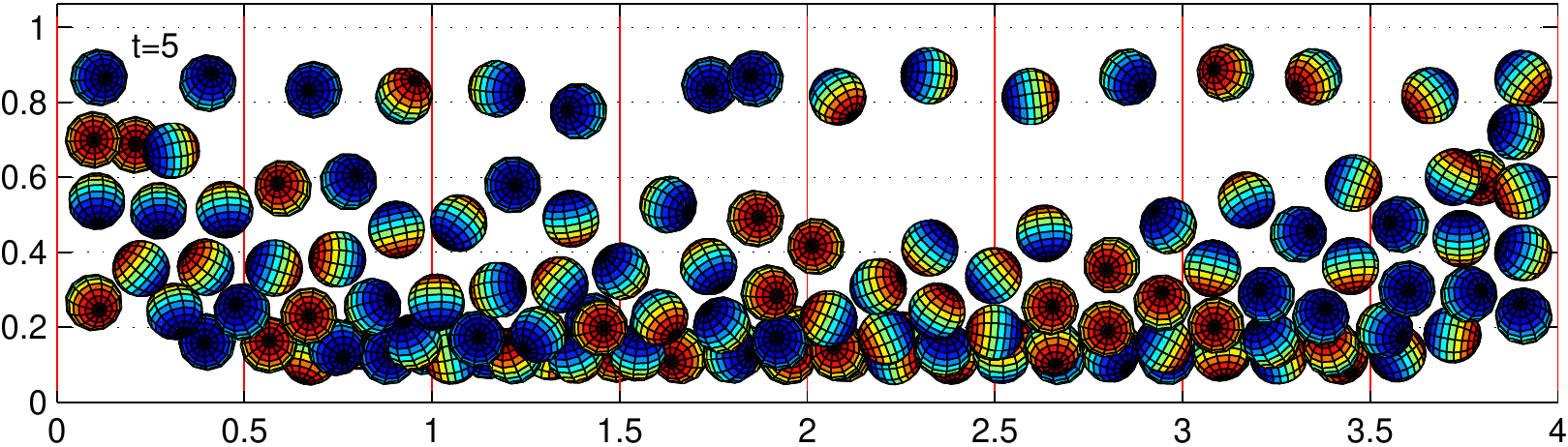}\\
\epsfxsize=2.9in
\epsffile{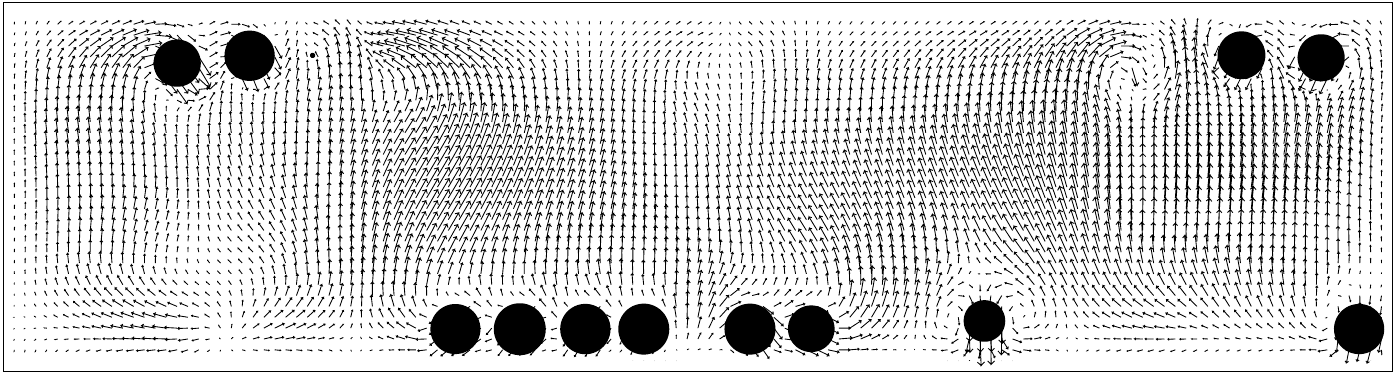}  \hskip 10pt
\epsfxsize=3.in
\epsffile{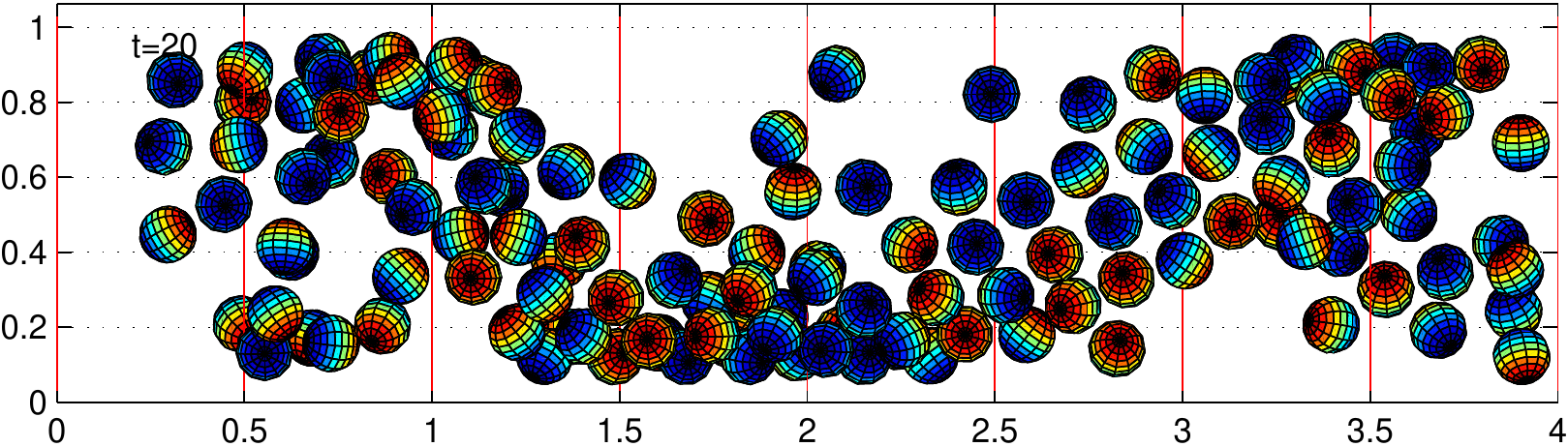}\\
\epsfxsize=2.9in
\epsffile{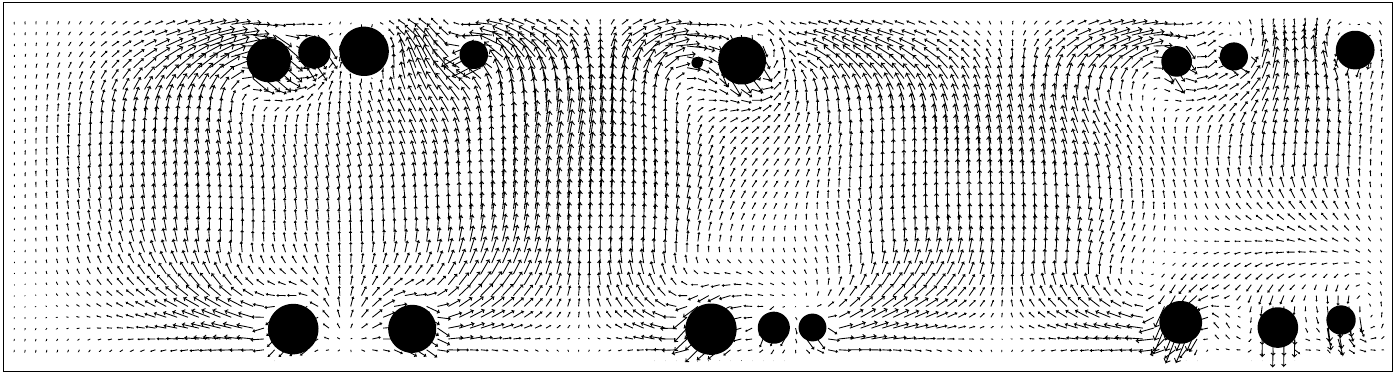}  \hskip 10pt
\epsfxsize=3.in
\epsffile{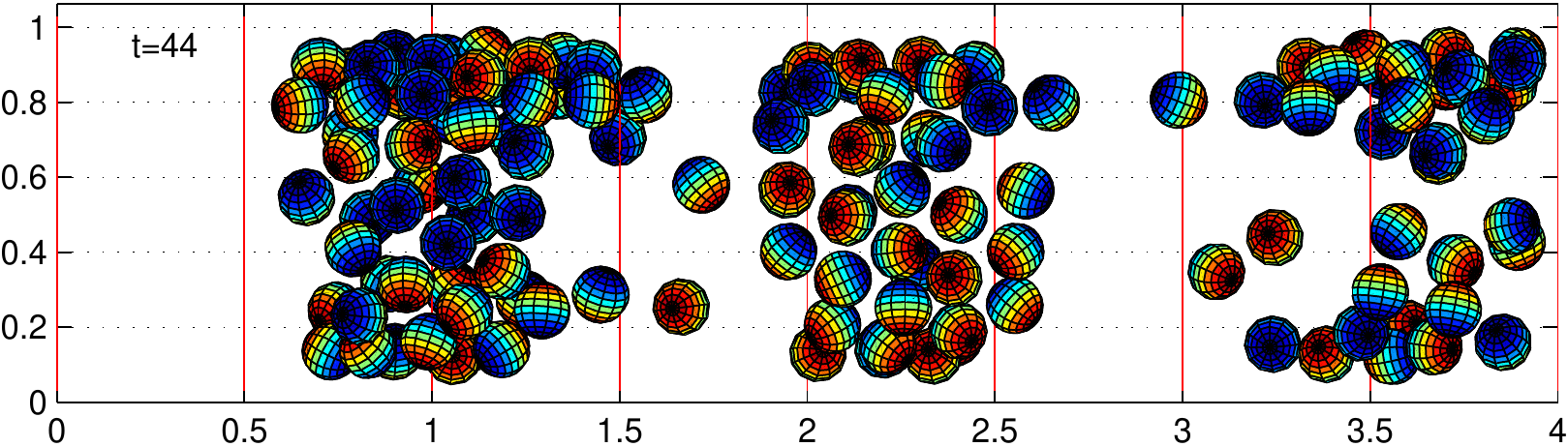}\\
\epsfxsize=2.9in
\epsffile{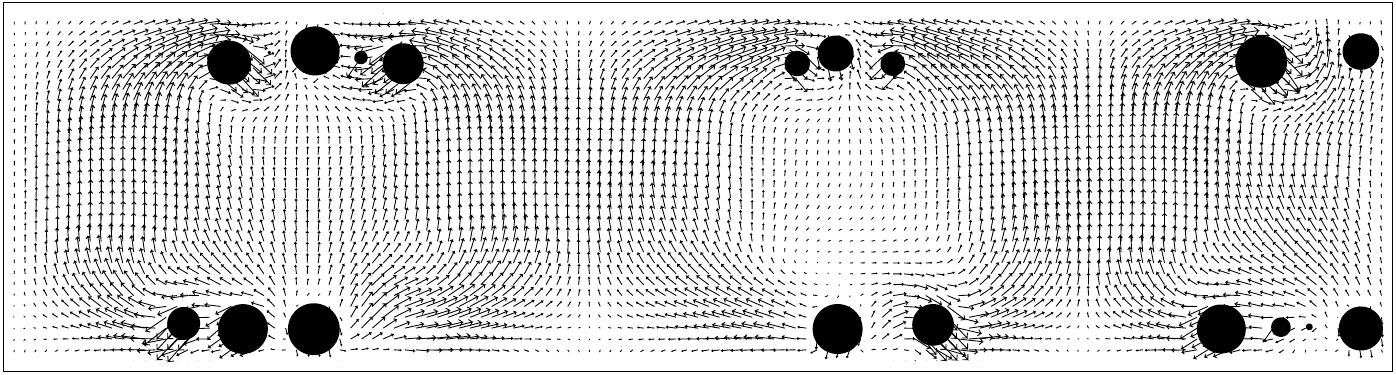}  \hskip 10pt
\epsfxsize=3.in
\epsffile{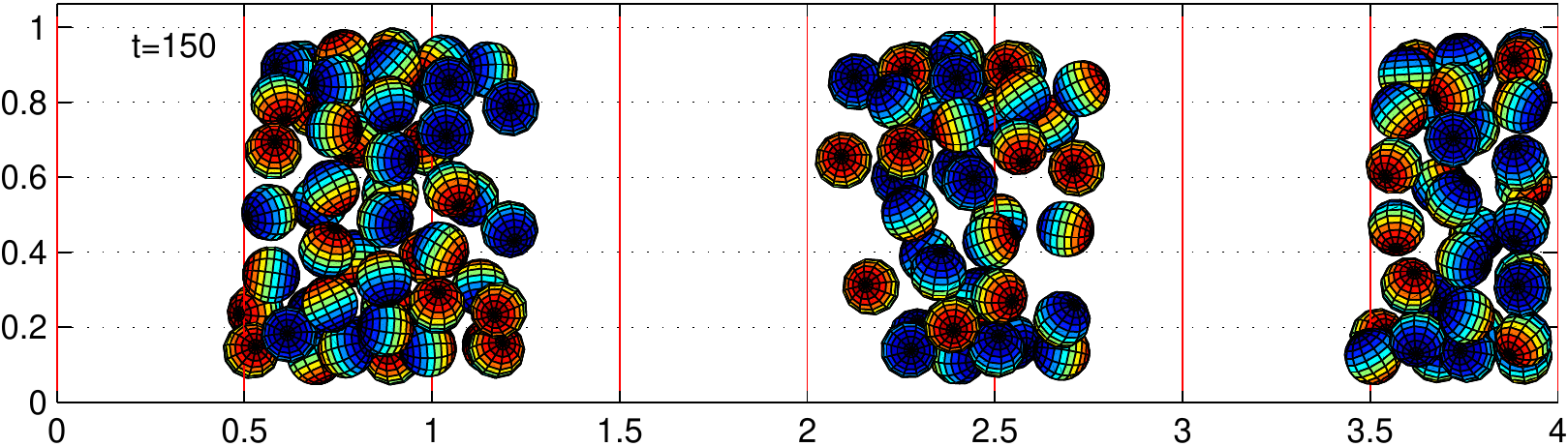}
\end{center}
\caption{(a) The projection of the velocity field on the vertical plane passing through the central
axis of the cylinder for the  case of 128 balls (left) and (b) the front view  of the position of 128 balls  (right) at $t=$ 2, 5, 44, 20 and 150 second (from top to bottom) with $\Omega=$ 12 sec$^{-1}$.}\label{fig.14}
\end{figure}  

To reproduce  the circular bands similar to the one in Figure 5 in \cite{Seiden2005}, we have considered the 
case of 128 balls in a truncated cylinder of length $L=$4 cm. We have first placed 128 balls  on 
16 circles in the middle of the cylinder with the initial gap size $d_g=a/4$ as in the previous 
subsection and then let them settle at the zero rotating rate. The particles are down to the bottom 
of the cylinder after 2 seconds as shown in Figure \ref{fig.14}. Then we rotate the cylinder at the 
rate $\Omega=$12 sec$^{-1}$. Those 128 particles first move up and down inside the rotating cylinder 
and interact with the fluid. At  $t=$ 5 second, there is no periodic flow field pattern in the cylinder 
axis direction. About $t=$ 20 second, two outer bands next to the two ends of the cylinder start forming.
Gradually three circular bands, which are similar to the one obtained experimentally in Figure 5 in \cite{Seiden2005}, 
are formed as shown in Figure \ref{fig.14}. All above cases show that the circular bands cause the flow 
field circulations, at least, for the cases considered in the paper.

\section{Conclusion}

In this article we have applied a distributed Lagrange multiplier fictitious domain method
to simulate rotating suspension of particles and to study the interaction between balls and 
fluid in a fully filled and horizontally rotating cylinder. The formation of circular bands 
studied in this paper is not resulted by by mutual interaction between the 
particles and the periodic inertial waves in the cylinder axis direction, but 
as the result of the interaction of particles.  When a circular band is forming, 
the part of the band formed by the particles moving downward becomes more compact 
due to the particle interaction strengthened by the downward acceleration from
the gravity. The  part of a band formed by the particles moving upward is always 
loosening up due to the slow down of the particle motion by the counter effect 
of the gravity.  To form a compact circular band (not a loosely one), enough particles 
are needed to have continuous interaction among themselves  through the entire circular band 
at a rotating rate 
so that the upward diffusion of particles can be overcome by the compactness process 
when these particles moving downward. Hence the balance of the gravity, the rotating rate, and 
the fluid flow inertia and the number of particles are important on the formation 
of circular bands.

\section*{Acknowledgments}
The authors acknowledge the helpful comments and suggestions of Howard H. Hu and Penger Tong. 
T.W. Pan and R. Glowinski acknowledge the support of NSF (grant DMS-0914788).

\end{document}